\documentclass[onecolumn]{emulateapj}
\usepackage{graphicx}
\begin{document}

\title{On the Mass Distribution and Birth Masses of Neutron Stars}

\author{Feryal \"Ozel$^1$, Dimitrios Psaltis$^1$, Ramesh Narayan$^2$, 
Antonio Santos Villarreal$^1$}

\affil{$^1$Department of Astronomy, University of Arizona, 933 N. 
Cherry Ave., Tucson, AZ 85721} 
\affil{$^2$ Harvard-Smithsonian Center for Astrophysics, 
60 Garden St., Cambridge, MA 02138} 

\begin{abstract}
We investigate the distribution of neutron star masses in different
populations of binaries, employing Bayesian statistical techniques.
In particular, we explore the differences in neutron star masses
between sources that have experienced distinct evolutionary paths and
accretion episodes. We find that the distribution of neutron star
masses in non-recycled eclipsing high-mass binaries as well as of slow
pulsars, which are all believed to be near their birth masses, has a
mean of $1.28~M_\odot$ and a dispersion of $0.24~M_\odot$. These
values are consistent with expectations for neutron star formation in
core-collapse supernovae. On the other hand, double neutron stars,
which are also believed to be near their birth masses, have a much
narrower mass distribution, peaking at $1.33~M_\odot$ but with a
dispersion of only $0.05~M_\odot$. Such a small dispersion cannot
easily be understood and perhaps points to a particular and rare
formation channel. The mass distribution of neutron stars that have
been recycled has a mean of $1.48~M_\odot$ and a dispersion of
$0.2~M_\odot$, consistent with the expectation that they have
experienced extended mass accretion episodes. The fact that only a
very small fraction of recycled neutron stars in the inferred
distribution have masses that exceed $\sim 2~M_\odot$ suggests that
only a few of these neutron stars cross the mass threshold to form low
mass black holes. 
\end{abstract}


\section{Introduction}

The mass distribution of neutron stars contains information about the
supernova explosion mechanisms, the equation of state of neutron star
matter, and the accretion history of each neutron star since its
formation. Certain populations of neutron stars such as those in
double neutron stars and in binaries with high mass companions are
thought to have experienced little-to-no accretion over their
lifetimes. In contrast, neutron stars in low-mass X-ray binaries and
fast pulsars, which are typically in close orbits around white dwarfs,
undergo extended accretion periods that are likely to move the neutron
star mass away from its birth value.

The neutron star mass measurements that were available a decade ago
allowed a statistical inference of the mass distribution of double
neutron stars (Finn 1994) or of pulsars in binaries, without
distinguishing between subgroups (Thorsett \& Chakrabarty 1999).  Finn
(1994) found that neutron star masses fall predominantly in the
$1.3-1.6~M_\odot$ range. Thorsett \& Chakrabarty (1999) found that the
mass distribution for the combined population is consistent with a
narrow Gaussian at $1.35 \pm 0.04~M_\odot$. More recently, Schwab,
Podsiadlowski, \& Rappaport (2010) argued that the distribution of
neutron star masses in double neutron stars is actually bimodal, with
one peak centered at $\sim 1.25~M_\odot$ and the other at $\sim
1.35~M_\odot$, which they attributed to different supernova explosion
mechanisms. Kiziltan, Kottas, \& Thorsett (2010), Valentim, Rangel, \&
Horvath (2011), and Zhang et al.\ (2011), on the other hand, inferred
the mass distribution of different neutron star subgroups based either
on the pulsar spin period or the binary companion, both of which were
taken to be indicative of the accretion history of the system. All
groups found that the neutron stars that are thought to have undergone
significant accretion are on average $0.2-0.3~M_\odot$ heavier than
those that have not.

One result that is common to all of these studies is the narrowness of
the mass distribution of double neutron stars, $\sigma \simeq
0.05~M_\odot$, which has been taken as indicative of the birth mass
distribution of all neutron stars. The mean of the distribution is at
$1.35~M_\odot$, which is significantly larger than the mass of the
pre-supernova iron core for neutron stars that form through the
core-collapse mechanism.  The Chandrasekhar mass for cores with
electron fractions in the range $Y_e = 0.42-0.48$ is
$1.15-1.34~M_\odot$. Electrostatic interactions and entropy of the
core introduce additional corrections to the pre-collapse mass (see
Timmes et al.\ 1996 for a discussion).  Taking into account the
binding energy of the neutron star results in gravitational masses for
the collapsed cores in the range $1.06-1.22~M_\odot$. Even the largest
of these masses is well below the mean of the observed distribution of
double neutron stars. Fallback of stellar matter onto the collapsing
core during the supernova explosion allows for the remnant to
increase. However, this is also expected to increase the dispersion of
masses by a comparable amount (see Zhang et al. 2008), which is
inconsistent with the narrowness of the inferred mass distribution of
double neutron star masses.

Considering a bimodal underlying distribution in the population of
double neutron stars, as in Schwab et al.\ (2010), makes the width of
each distribution even narrower: $0.008~M_\odot$ and $0.025~M_\odot$
for the two components. For the lower mass component centered around
$\sim 1.25~M_\odot$, such a narrow distribution may be reasonably
obtained through an electron capture supernova, the onset of which
occurs at a particular mass threshold of an ONeMg white dwarf
(Podsiadlowski et al.\ 2005). However, the second component, which is
centered at $1.35~M_\odot$ cannot be explained as a result of the
electron capture supernovae and poses the same challenge in its
narrowness when explained via the core-collapse mechanism.

In order to model the distribution of neutron star masses both at
their births and throughout their lives, one important question to
address is whether double neutron stars are a representative sample
for neutron stars at their birth masses. In this paper, we address
this question by identifying a different population of neutron stars
at or near their birth masses and compare the inferred mass
distribution with that of double neutron stars. Furthermore, to
pinpoint the effects of subsequent accretion, we compare the inferred
mass distribution of these neutron stars to that of neutron stars
which have undergone or are currently undergoing extended mass
accretion.

Making use of all of the currently available neutron star mass
measurements, we divide the sample into various subgroups based on the
nature of the companion as well as the neutron star spin. We employ a
uniform Bayesian statistical approach that utilizes the entire
posterior likelihood of each mass measurement to infer the parameters
of the underlying mass distribution without assuming Gaussian errors.

In Section 2, we present the neutron star mass measurements, grouping
the data according to the measurement technique and the type of the
binary. In Section 3, we estimate the parameters of the underlying
mass distribution for each subgroup and assess the sensitivity of our
results on the particular measurements and priors. In section 4, we
summarize our findings and discuss their implications for the
different physical mechanisms that determine the neutron star mass
distribution. 

\section{Mass Measurements}

Mass measurements of neutron stars are carried out in several
different ways. When neutron stars appear as pulsars, high precision
pulse timing observations lead to a measurement of several orbital
parameters.  The pulsar's orbit can be described in classical gravity
by the five Keplerian parameters: the binary period $P_b$, the
eccentricity $e$, the projection of the pulsar's semi-major axis on
the observer's line of sight $a \sin i$, the time of periastron $T_0$,
and the longitude of periastron $\omega_0$, where $i$ is the angle
between the orbital angular momentum vector and the line of sight. The
mass function, which is related to the mass of the pulsar $M_{\rm psr}$,
its companion $M_{\rm c}$, and the inclination angle $i$, 
\begin{equation}
f = \frac{(M_{\rm c} \sin i)^3}{M_{\rm T}^2} = \left(\frac{2 \pi}
{P_b}\right)^2 \frac{(a \sin i)^3}{G},  
\end{equation}
is, therefore, directly obtained from these orbital parameters, where
$M_{\rm T} = M_{\rm psr} + M_{\rm c}$ is the total mass of the system.

Proceeding from a mass function to a measurement of the mass of the
pulsar and of its companion requires additional information. This
information can come from measurement of relativistic effects in the
binary orbits or from independent observations of the companion
stars. We discuss below the various techniques, the measurements they
resulted in to date, and the associated uncertainties.

The majority of the precise neutron star mass measurements come from
radio pulsar timing techniques and rely on the measurement of
relativistic effects in the binary orbits.

The general relativistic effects can be described by five additional
``post-Keplerian'' (or PK) parameters.  These are: the advance of
periastron $\dot{\omega}$, the orbital period decay $\dot{P}_{b}$, the
time dilation-gravitational redshift factor $\gamma$, as well as the
range $r$ and the shape $s$ of Shapiro delay, which are related to the
component masses, the orbital period, and eccentricity by
\begin{equation}
\dot{\omega} =  3\left(\frac{P_{b}}{2\pi}\right)^{-5/3}
\left( \frac{G M_{\rm T}}{c^3} \right)^{2/3}\left(1-e^2\right)^{-1},
\end{equation}
\begin{equation}
\dot{P}_b = -\frac{192\pi}{5}\left(\frac{P_{b}}{2\pi}\right)^{-5/3}  
\left(\frac{G}{c^3}\right)^{5/3}  
\left(1+\frac{73}{24}e^2+\frac{37}{96}e^4\right)\times \nonumber\\
(1-e^2)^{-7/2}\, M_{\rm psr} M_{\rm c}\,M_{\rm T}^{-1/3} ,
\end{equation}
\begin{equation}
\gamma = e\left(\frac{P_{b}}{2\pi}\right)^{1/3}
\left(\frac{G}{c^3}\right)^{2/3} M_{\rm T}^{-4/3}
M_{\rm c}\left(M_{\rm psr} + 2 M_{\rm c} \right) ,
\end{equation}
\begin{equation}
r = \frac{G}{c^3} M_{\rm c},
\end{equation}
\begin{equation}
s = G^{-1/3} a\sin i \left(\frac{P_{b}}{2\pi}\right)^{-2/3}
M_{\rm T}^{2/3}\, M_{\rm c}^{-1}.
\label{Eq:PK}
\end{equation}

In highly eccentric systems that have been observed repeatedly over a
long period of time, the measurement of $\dot{\omega}$ is usually
possible and leads to a strong constraint on the total mass of the
binary $M_{\rm T}$. The measurement of the parameter $\gamma$ requires
similarly eccentric systems and long-term monitoring. In high
inclination systems, on the other hand, it is sometimes possible to
detect the Shapiro time delay and obtain the parameters $r$ and
$s$. The measurement of the rate of orbital period decay typically
requires the longest monitoring and timing of the pulsar, sometimes
over decades. This has been achieved for a handful of pulsars in
binaries.

The precision with which the pulsar mass can be determined ultimately
depends on the number of PK parameters that are measured for that
binary. In systems where two or more PK parameters are known, the
pulsar mass is precisely determined. In the category where only one PK
parameter is known in addition to the mass function, the mass of each
neutron star is not as well constrained.

In the following, we divide neutron star mass measurements into
categories based on the information available for each binary system
such as the number of PK parameters. For each category, we derive the
likelihood $P_i({\rm data} \vert M_{\rm NS})$, which measures the
chance of obtaining the particular set of data for the $i$-th source
if that source had mass $M_{\rm NS}$.

We are ultimately interested in delineating the effects of mass
accretion from the neutron star birth masses. We, therefore, further
divide each category into groups based on the nature of the companion
star or the spin of the pulsar.  In particular, neutron star-neutron
star (NS-NS) binaries, as well as eclipsing X-ray pulsars in high-mass
X-ray binaries and slow radio pulsars are expected to have experienced
little-to-no accretion. On the other hand, neutron star-white dwarf
(NS-WD) binaries are the remnants of a long-lasting low-mass X-ray
binary phase, where significant mass accretion may have occurred. We
also group millisecond pulsars with main sequence companions (NS-MS)
along with the latter group, because of the probable recycling these
neutron stars underwent to reach millisecond periods. We will refer to
the latter group as ``fast pulsars''. Finally, we will also consider
accreting X-ray bursters, for which the masses have been measured
primarily through X-ray spectroscopy.

In detail, the various categories are:

{\it (Ia) Double Neutron Stars with at least 2 PK Parameters.---\/}
Six neutron star systems shown in Table~1 have at least 2 measured PK
parameters leading to well determined masses. In this case, the
likelihood of neutron star masses is highly symmetric and narrowly
peaked, and thus can be described as a Gaussian
\begin{equation}
P_i({\rm data} \vert M_{\rm NS}) = C_i \exp\left[-\frac{(M_{\rm NS}-M_{0,i})^2}
{2\sigma_{M,i}^2} \right]
\end{equation}
with a mean $M_{0,i}$ and a standard deviation $\sigma_{M,i}$. In this
and the following expressions, $C_i$ is a proper normalization
constant such that
\begin{equation}
\int_0^\infty P_i({\rm data} \vert M_{\rm NS}) dM_{\rm NS} = 1.
\end{equation} 


\begin{deluxetable}{llll}
\tabletypesize{\scriptsize}
\tablewidth{300pt}
\tablenum{1}
\tablecaption{Precise Masses of Double Neutron Star Systems\tablenotemark{a}}
\tablehead{
 \colhead{Name} &
 \colhead{Mass} &
 \colhead{Error} &
 \colhead{Refs\tablenotemark{b}} 
 \cr
 \colhead{} &
 \colhead{($M_{\odot}$)} &
 \colhead{($M_{\odot}$)} &
 \colhead{}
}
\startdata
%
J0737-3039&	 $1.3381$&       $0.0007$&	 1\\
\hspace*{0.4cm}
pulsar B&        $1.2489$&       $0.0007$&	 1\\ 
B1534+12&	 $1.3332$&   	 $0.0010$&	 2\\
\hspace*{0.4cm}
companion&       $1.3452$&       $0.0010$&	 2\\ 
J1756-2251&      $1.312$&        $0.017$&	 3\\ 
\hspace*{0.4cm}
companion&       $1.258$&        $0.018$&	 3\\  
J1906+0746&	 $1.323$&	 $0.011$&	 4, 5\\
\hspace*{0.4cm}
companion&       $1.290$&        $0.011$&	 4, 5\\ 
B1913+16&	 $1.4398$& 	 $0.002$&	 6\\
\hspace*{0.4cm}
companion&       $1.3886$&       $0.002$&	 6\\ 
B2127+11C& 	 $1.358$&        $0.010$&	 7\\  
\hspace*{0.4cm}
companion&	 $1.354$&        $0.010$&	 7
\enddata
\scriptsize
\tablenotetext{a} {Defined as systems with $\ge 2$ PK parameters measured.}
\tablenotetext{b} {References: 
1.\ Kramer et al.\ 2006;
2.\ Stairs et al.\ 2002;
3.\ Ferdman 2008;
4.\ Lorimer et al.\ 2006; 
5.\ Kasian 2012;
6.\ Weisberg et al. 2010; 
7.\ Jacoby et al. 2006 \\
}
\mbox{}
\end{deluxetable}

\begin{figure}
\centering
   \includegraphics[scale=0.75]{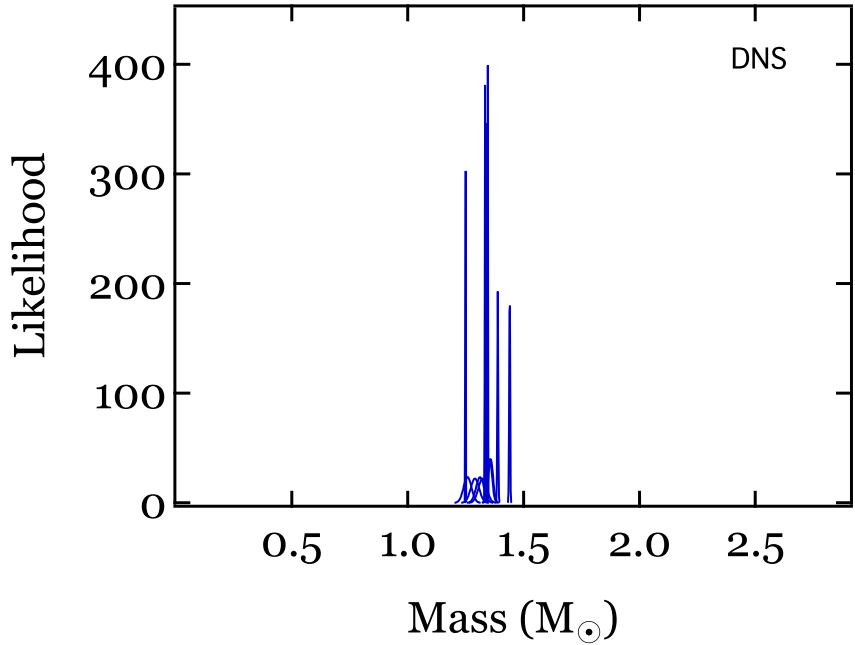}
\caption{The likelihoods $P_i({\rm data}\vert M_{\rm NS})$ for the 12
double neutron stars with precisely determined masses arising from the
measurement of $\ge 2$ PK parameters. These systems belong to category
Ia discussed in the text.}
\mbox{}
\label{fig:psr_gauss} 
\end{figure}

We plot in Figure~\ref{fig:psr_gauss} the likelihood of the masses
of NS-NS binaries that belong to this category.

\begin{figure}
\centering
   \includegraphics[scale=0.75]{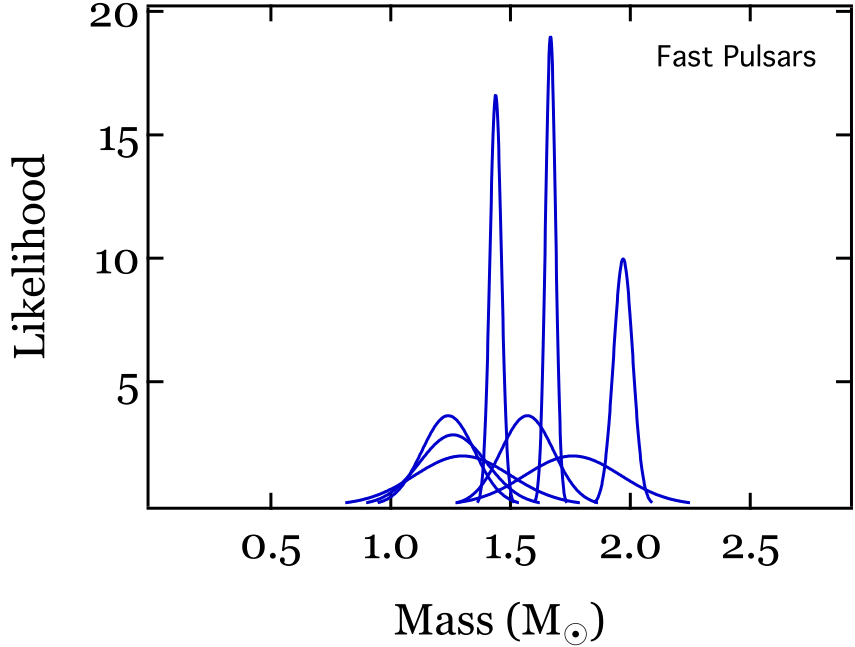}
\caption{The likelihoods $P_i({\rm data}\vert M_{\rm psr})$ for nine
recycled pulsars with white dwarf companions. The measurement of $\ge
2$ PK parameters in these binaries lead to precisely determined
neutron star masses. These systems belong to category Ib discussed in
the text.}
\mbox{}
\label{fig:psr_recycl_gauss} 
\end{figure}

{\it (Ib) Neutron Stars in Binaries with at least 2 PK
Parameters.---\/} To date, observations of nine neutron stars that are
in close orbits around predominantly white dwarf companions have
yielded a measurement of at least 2 PK parameters. The known exception
is PSR~J1903+0327, with a probable main sequence companion, but is
nevertheless thought to be recycled owing to its millisecond
period. As in the previous category, we assign a Gaussian likelihood
to each measurement with a mean $M_{0,i}$ and a standard deviation
$\sigma_{M,i}$. We present in Table~2 and
Figure~\ref{fig:psr_recycl_gauss} the mass measurements and their
uncertainties for these NS-WD binaries.  Even though it has a white
dwarf companion, PSR J1141$-$6545 is different from the rest of the
sources in this category in that it is a slowly spinning neutron
star. For this reason, we group it with the accreting and slow pulsars
discussed below and show its likelihood in Figure~\ref{fig:psr_slow}.


\begin{deluxetable}{llll}
\tabletypesize{\scriptsize}
\tablewidth{300pt}
\tablenum{2}
\tablecaption{Precise Masses of Neutron Stars with White Dwarf 
Companions\tablenotemark{a}}
\tablehead{
 \colhead{Name} &
 \colhead{Mass} &
 \colhead{Error} &
 \colhead{Refs\tablenotemark{b}} 
 \cr
 \colhead{} &
 \colhead{($M_{\odot}$)} &
 \colhead{($M_{\odot}$)} &
 \colhead{}
}
\startdata
%
J0437-4715&	 $1.76$&         $0.2$&	 	 1\\
J0751+1807&      $1.26$&         $0.14$&	 2, 3\\ 
J1141-6545&	 $1.27$&   	 $0.01$&	 4\\
J1614-2230&      $1.97$&         $0.04$&	 5\\ 
J1713+0747&	 $1.30$&	 $0.2$&	         6\\
J1802-2124&      $1.24$&         $0.11$&	 7\\ 
B1855+09&        $1.57$&         $0.11$&	 8, 9\\ 
J1903+0327&      $1.667$&        $0.021$&        10\\
J1909-3744&      $1.438$&        $0.024$&	 11
\enddata
\scriptsize
\tablenotetext{a} {Defined as systems with $\ge 2$ PK parameters measured.}
\tablenotetext{b} {References: 
1.\ Verbiest et al. 2008;
2.\ Nice et al.\ 2005;
3.\ Nice et al.\ 2008;
4.\ Bhat et al.\ 2008;
5.\ Demorest et al.\ 2010;
6.\ Splaver et al.\ 2005;
7.\ Ferdman et al.\ 2010;
8.\ Nice et al.\ 2003; 
9.\ Kaspi et al.\ 1994; 
10.\ Freire et al.\ 2011; 
11.\ Jacoby et al.\ 2005 \\
}
\mbox{}
\end{deluxetable}


{\it (IIa) Double Neutron Stars with 1 PK Parameter.---\/} 
In this category, there are three double neutron stars, for which 
the measurement of the advance of periastron $\dot{\omega}$
allows for a precise determination of the total mass of the binary.
However, in the absence of a second PK parameter or a knowledge of the
system inclination $i$, the mass of each neutron star is not well
constrained. In these cases, using the total mass of the system as 
a constraint allows us to write the likelihood of the mass of the 
pulsar as
\begin{equation} 
P_i({\rm data} \vert M_{\rm psr}) = C_i \int dM_{\rm tot} 
\exp\left[-\frac{(M_{\rm tot}-M_{{\rm tot},0})^2}
{2\sigma_{M_{\rm tot},i}^2} \right] \int d(\cos i)
\exp\left[-\frac{\left(f_0-\frac{M^3_{\rm psr}\sin^3i}{M^2_{\rm tot}}\right)^2}
{2\sigma_f^2}\right]
\label{eq:pm_omegadot}
\end{equation}
and of the companion as 
\begin{equation} 
P_i({\rm data} \vert M_{\rm c}) = C_i \int dM_{\rm tot} 
\exp\left[-\frac{(M_{\rm tot}-M_{{\rm tot},0})^2}
{2\sigma_{M_{\rm tot},i}^2} \right] \int d(\cos i)
\exp\left[-\frac{\left(f_0-\frac{(M_{\rm tot}-M_{\rm c})^3
\sin^3i}{M^2_{\rm tot}}\right)^2}
{2\sigma_f^2}\right]. 
\label{eq:pm__c_omegadot}
\end{equation}

\begin{figure}
\centering
   \includegraphics[scale=0.75]{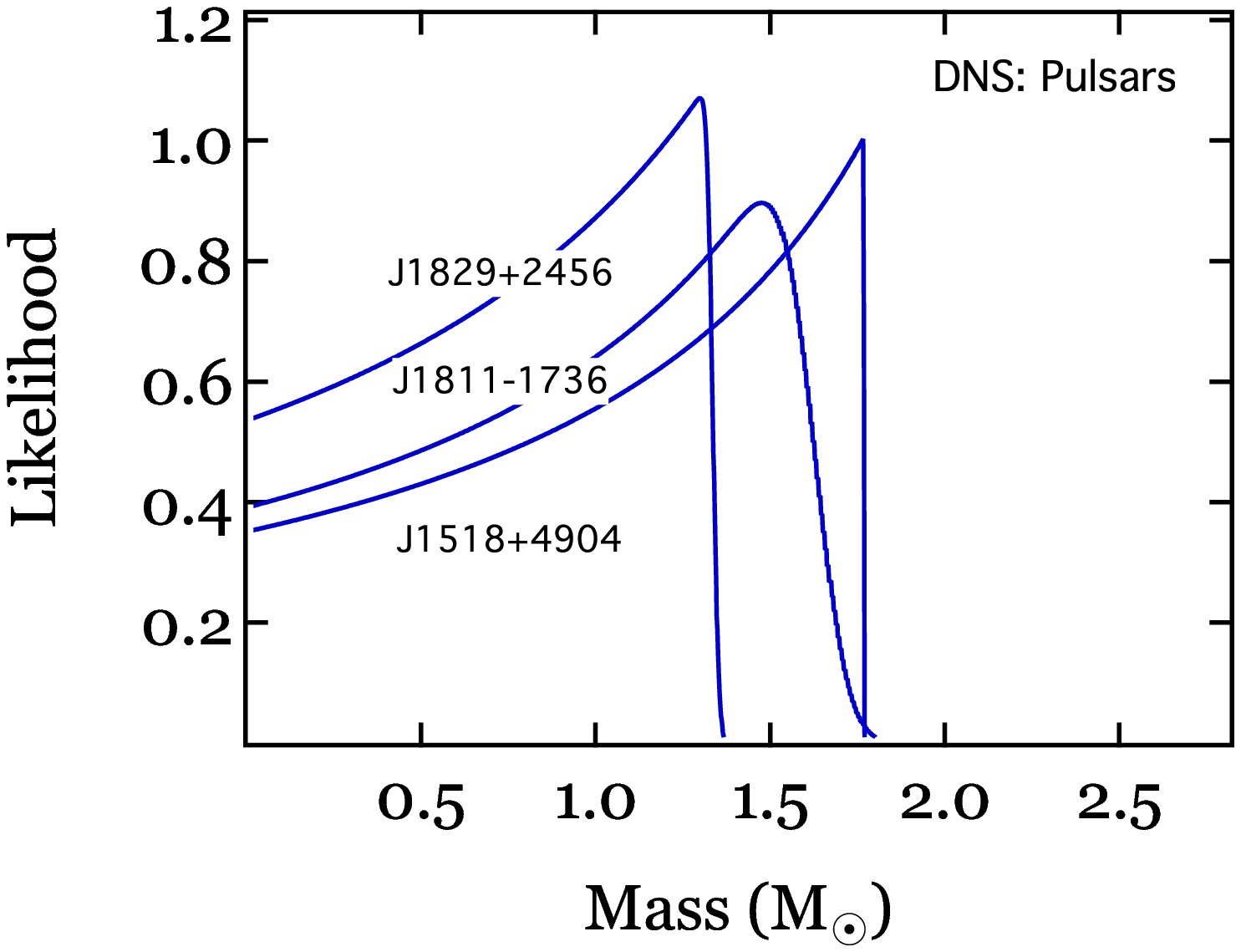}
   \includegraphics[scale=0.75]{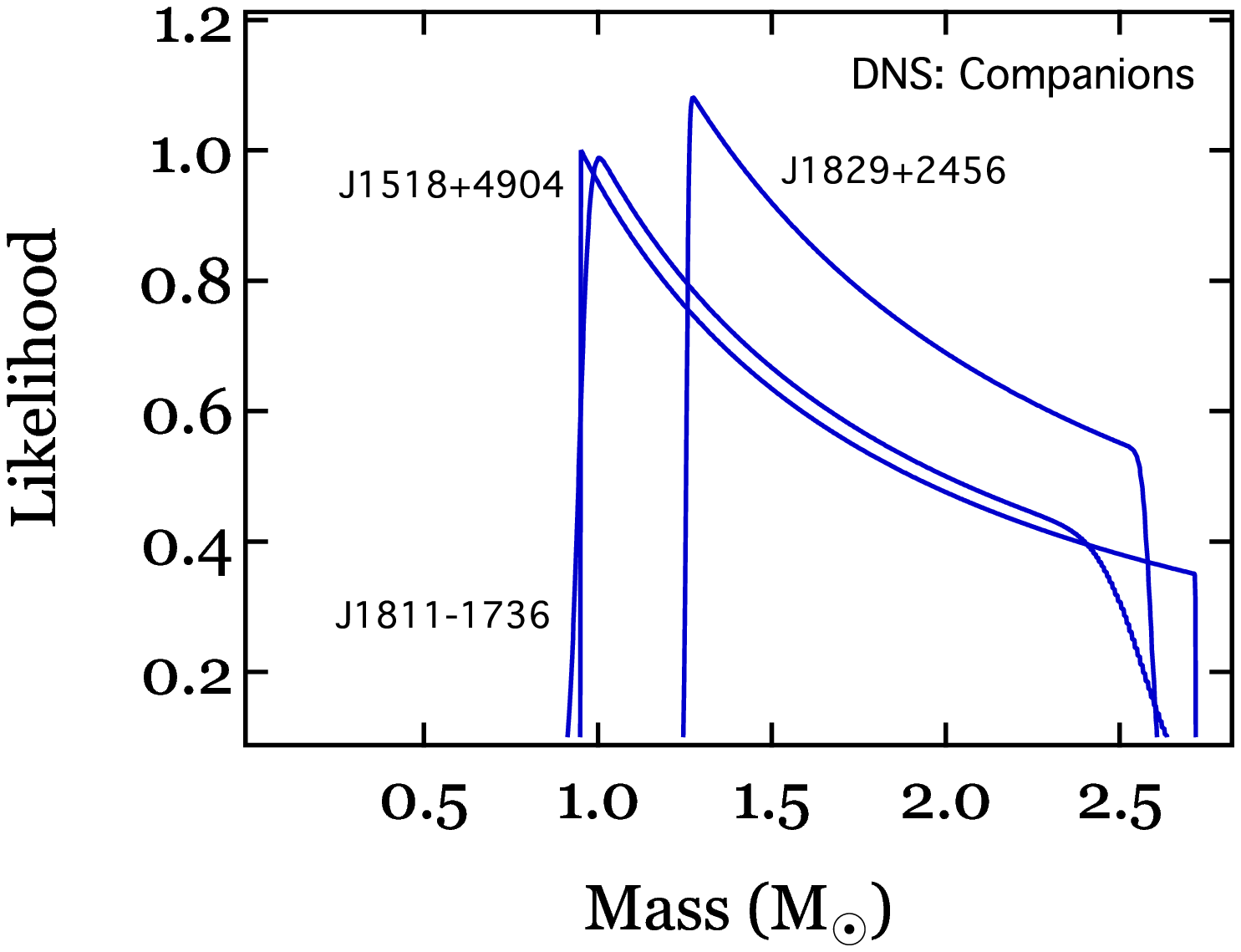}
\caption{The likelihoods $P_i({\rm data}\vert M_{\rm NS})$ for the
double neutron stars with 1 PK parameter for (top) the pulsars and
(bottom) companion neutron stars. These belong to category IIa
discussed in the text.}
\mbox{}
\label{fig:psr_omega} 
\end{figure}

We present in Table~3 the relevant pulsar data for the three double
neutron star binaries that are in this category.
Figure~\ref{fig:psr_omega} shows the likelihood of each neutron star
mass $P_i({\rm data}\vert M_{\rm NS})$ for these sources, with the top
panel including the pulsars and the bottom panel the companions. Note,
however, that even though they are shown independently for the
purposes of this figure, the likelihoods of the masses of the pulsar
and its companion are not independent probabilities. Therefore, when
inferring the mass distribution of double neutron stars, the
constraint over the total mass is incorporated as we will discuss in
\S3.


\begin{deluxetable}{llllllll}
\tabletypesize{\scriptsize}
\tablewidth{300pt}
\tablenum{3}
\tablecaption{Dynamical Data for Double Neutron Stars with 1 PK parameter}
\tablehead{
 \colhead{Name} & \colhead{$f(M)$} & \colhead{$\dot{\omega}$} &
 \colhead{$M_{tot}$} & \colhead{Refs\tablenotemark{a}} \cr \colhead{}
 & \colhead{($M_{\odot}$)} & \colhead{(deg yr$^{-1}$)} &
 \colhead{($M_{\odot}$)} & \colhead{} }
\startdata
%
PSR J1518+4904 &   0.115988 &      0.0113725(9) &  2.7183(7) &	      1 \\
PSR J1811-1736 &   0.128121(5) &   0.0090(2) &     2.57(10)  &        2 \\
PSR J1829+2456 &   0.29413(1) &    0.2929(16) &    2.59(2)   &        3
\enddata
\scriptsize
\tablenotetext{a} {References: 
1.\ Janssen et al. 2008;
2.\ Corongiu et al. 2007;
3.\ Champion et al. 2005 \\
}
\mbox{}
\end{deluxetable}


\begin{figure}
\centering
   \includegraphics[scale=0.75]{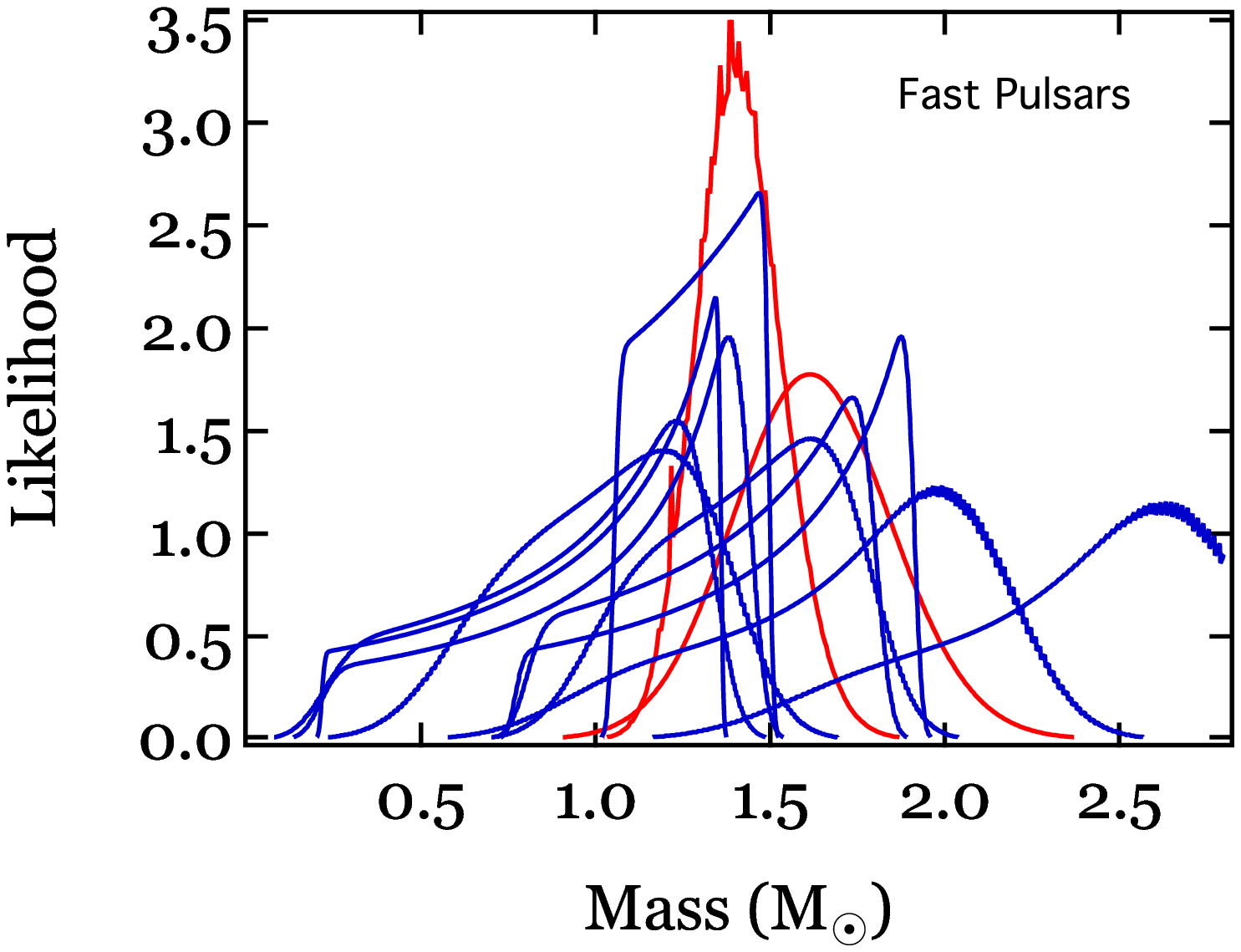}
\caption{The likelihoods $P_i({\rm data}\vert M)$ for 
the recycled neutron stars with 1 PK parameter (blue curves; category
IIb) or with optical observations of the white dwarf companions (red
curves; category III).  }
\mbox{}
\label{fig:psr_recycl_omega} 
\end{figure}

\begin{figure}
\centering
   \includegraphics[scale=0.75]{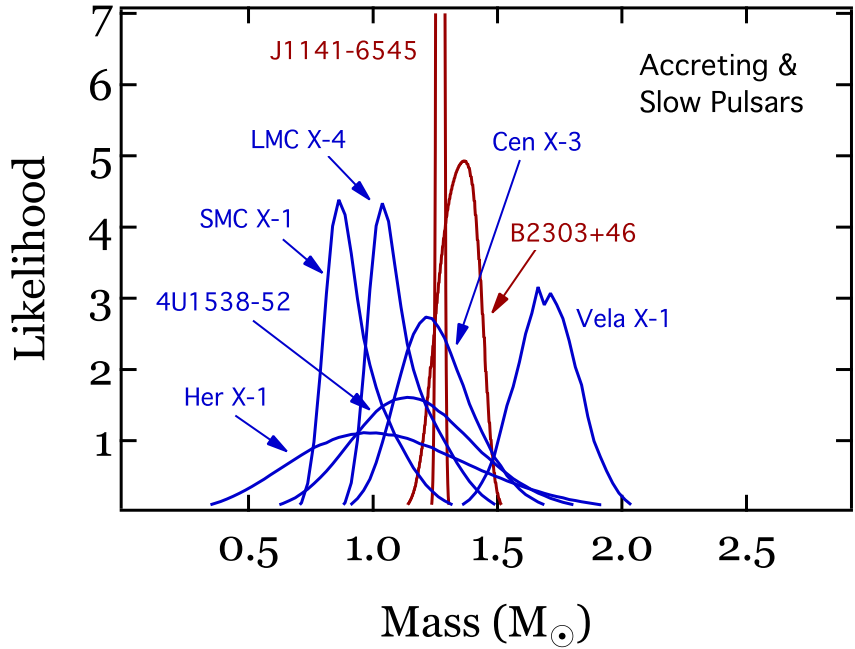}
\caption{The likelihoods $P_i({\rm data}\vert M_{\rm NS})$ for 
the eclipsing X-ray pulsars in high-mass X-ray binaries, which belong
to category IV discussed in the text. This figure also includes the
likelihoods for the slow pulsars PSR~J1141$-$6545 and PSR~B2303+46,
which belong to categories Ib and IIb, respectively, but have not been
recycled.}
\mbox{}
\label{fig:psr_slow} 
\end{figure}

{\it (IIb) Neutron Stars in Binaries with 1 PK Parameter.---\/} This
category comprises of eleven neutron star binaries with mostly white
dwarf companions. In several systems, the companions have not been
identified. A constraint on the total mass comes from the measurement
of the rate of advance of the periastron. We assign to each neutron
star mass a likelihood according to equation~(\ref{eq:pm_omegadot}).
Table~4 and Figure~\ref{fig:psr_recycl_omega} shows the relevant
parameters for these systems. Note that, of this category, PSR B2303+46
is a slowly spinning neutron star. For this reason, we group it with
the accreting and slow pulsars discussed below and show its likelihood
in Figure~\ref{fig:psr_slow}.


\begin{deluxetable}{llllllll}
\tabletypesize{\scriptsize}
\tablewidth{300pt}
\tablenum{4}
\tablecaption{Data for NS-WD binaries with 1 PK parameter}
\tablehead{
 \colhead{Name} & \colhead{$f(M)$} & \colhead{$\dot{\omega}$} &
 \colhead{$M_{tot}$} & \colhead{Refs\tablenotemark{a}} \cr \colhead{}
 & \colhead{($M_{\odot}$)} & \colhead{(deg yr$^{-1}$)} &
 \colhead{($M_{\odot}$)} & \colhead{} }
\startdata
%
J0024-7204H & 0.001927     &    0.066(2)    & 1.61(4)   &  1 \\
J0514-4002A & 0.14549547   &    0.01289(4)  & 2.453(14) &  2 \\
J0621+1002  & 0.027026849  &    0.0102(2)   & 2.32(8)   &  3 \\ 
B1516+02B   & 0.000646723  &    0.0142(7)   & 2.29(17)  &  4 \\
J1748-2021B & 0.0002266235 &    0.00391(18) & 2.92(20)  &  5 \\
J1748-2446I & 0.003658     &                & 2.17(2)   &  4, 6 \\
J1748-2446J & 0.013066     &                & 2.20(4)   &  4, 6 \\
J1750-37A   & 0.0518649    &    0.00548(30) & 1.97(15)  &  5 \\
B1802-07    & 0.00945034   &    0.0578(16)  & 1.62(7)   &  7 \\
J1824-2452C & 0.006553     &                & 1.616(7)  &  4 \\
B2303+46    & 0.246332	   &    0.01019(13) & 2.64(5)   &  7
\enddata
\scriptsize
\tablenotetext{a} {References: 
1.\ Freire et al.\ 2003;
2.\ Freire et al.\ 2007; 
3.\ Kasian 2012;
4.\ Freire et al.\ 2008a; 
5.\ Freire et al.\ 2008b;
6.\ Ransom et al.\ 2005; 
7.\ Thorsett \& Chakrabarty 1999  \\
}
\mbox{}
\end{deluxetable}


{\it (III) Neutron Stars in Binaries with Optical Observations of
White Dwarf Companions.---\/} For two neutron stars in orbit around
white dwarfs, optical observations of the companions have resulted in
the measurements of the mass $M_{\rm WD}$ as well as of the radial
velocity amplitude $K_{\rm opt}$ of the white dwarf. The latter, in
combination with the orbital parameters obtained from the radio timing
solution gives the mass ratio of the binary according to
\begin{equation}
q = K_{\rm opt} P_{\rm b} \frac{(1-e^2)^{1/2}}{2 \pi a_{\rm psr} \sin i}.
\end{equation}
Using this information, we can then calculate the likelihood of the 
neutron star mass as 
\begin{equation}
P_i({\rm data} \vert M_{\rm psr}) = C_i \int dM_{\rm WD} \int di \sin i
\exp\left[-\frac{(M_{\rm WD}-M_{{\rm WD},0})^2}
{2\sigma_{M_{\rm WD},i}^2} \right] 
\exp\left[-\frac{(\frac{M_{\rm psr}}{M_{\rm WD}}-q_{i,0})^2}
{2\sigma_{q,i}^2} \right] \delta[f(i)-f_{0,i}],
\end{equation}
where we assume that the error in the measurement of each mass function 
is negligible. We perform the integration over inclination making use of 
the identity
\begin{equation}
\delta[f(i)-f_{0,i}] = \frac{\delta(i-i_0)}{\vert df/di\vert_{i_0}}
\end{equation}
where $i_0$ is the solution of the equation $f(i_0) - f_{0,i}=0$, i.e.,
\begin{equation}
\sin i_0 = \left[f_{0,i} \left(1+\frac{M_{\rm psr}}{M_{\rm WD}}\right)^2 
M_{\rm WD}^{-1}\right]^{1/3}.
\end{equation}
Given that 
\begin{equation}
\left\vert \frac{df}{di}\right\vert_{i_0} = 3 f \left\vert 
\frac{\cos i_0}{\sin i_0}\right\vert, 
\end{equation}
the likelihood becomes
\begin{equation}
P_i({\rm data} \vert M_{\rm psr}) = \frac{C_i}{3 f_{0,i}} 
\int dM_{\rm WD} \frac{\sin^2 i_0}{\cos i_0}
\exp\left[-\frac{(M_{\rm WD}-M_{{\rm WD},0})^2}
{2\sigma_{M_{\rm WD},i}^2} \right]
\exp\left[-\frac{(\frac{M_{\rm psr}}{M_{\rm WD}}-q_{i,0})^2}
{2\sigma_{q,i}^2} \right]\;.
\end{equation}

Table~5 summarizes the relevant data for these two binaries. The
likelihoods of the neutron star masses are shown as red curves in
Figure~\ref{fig:psr_recycl_omega}.


\begin{deluxetable}{lllll}
\tabletypesize{\scriptsize}
\tablewidth{300pt}
\tablenum{5}
\tablecaption{Data for NS-WD Binaries with Optical Observations}
\tablehead{
 \colhead{Name} & \colhead{$f(M)$} & \colhead{$M_{\rm WD}$} &
\colhead{$q$} & \colhead{Refs\tablenotemark{a}} 
\cr 
\colhead{} & \colhead{($M_{\odot}$)} & \colhead{($M_{\odot}$)} & 
\colhead{} & \colhead{} 
}
\startdata
%
J1012+5307  &  0.00058709(2)   &  0.156$\pm$0.02  &  10.7$\pm$0.5   & 1,2 \\
B1911-5958A &  0.002687603(13) &  0.18$\pm$0.02   &   7.36$\pm$0.25 & 3,4 
\enddata
\scriptsize
\tablenotetext{a} {References: 
1.\ Callanan et al. 1998; 
2.\ Nicastro et al. 1995; 
3.\ Bassa et al. 2006; 
4.\ D'Amico et al. 2002 \\
}
\mbox{}
\end{deluxetable}

{\it (IV) Eclipsing X-ray Pulsars.---\/} Eclipsing X-ray pulsars
provide a wealth of observational information using which we can
estimate the masses of these neutron stars.  X-ray observations of
each pulsar give the orbital period of the binary $P_b$, the
eccentricity of the orbit $e$, longitude of periastron $\omega_0$, the
semi-major axis of the neutron star's orbit $a_X\sin i$, and the
semi-duration of the eclipse $\theta_e$.  In addition, optical
observations of the companion star give its velocity amplitude $K_{\rm
opt}$, its projected rotational velocity $v_{\rm rot} \sin i$, and the
amplitude of ellipsoidal variations $A$. From these observables, it is
possible to solve for the fundamental parameters of the binary, viz.,
the mass of the neutron star $M_{\rm NS}$, the mass, radius and
rotational angular velocity of the companion, $M_{\rm opt}$, $R_{\rm
opt}$, $\Omega_{\rm opt}$, and the inclination angle of the binary
$i$.

The necessary data are available for six eclipsing pulsars: Vela X--1,
4U1538--52, SMC X--1, LMC X--4, Cen X--3, and Her X--1 (see, e.g., van
Kerkwijk, van Paradijs, \& Zuiderwijk 1995; van der Meer et
al. 2007). Recently, Rawls et al. (2011) collected all the available
data and presented a detailed analysis of the likelihood of mass for
the individual neutron stars; see Table~6 for a compilation of the
relevant results. 

There are several sources of potential systematic uncertainties in the
masses inferred in these eclipsing binaries. For example, there are
significant residuals in the radial velocity curves of Vela X-1 after
the best-fit orbital solution is subtracted (e.g., Barziv et
al. 2001), in which Quaintrell et al. (2003) noted the presence of a
periodicity and suggested modes on the star as a possible origin. More
recently, Koenigsberger et al.\ (2012) developed a model for these
residuals based on the interaction between the neutron star and its
companion but did not directly fit the model to the data. Even though
they concluded that a 1.55~$M_\odot$ neutron star is marginally
consistent with observations, a higher mass for the neutron star,
centered around 1.7~$M_\odot$, appears to be favored.

A second important source of systematic uncertainty arises from
modeling of the ellipsoidal modulations, which includes a contribution
from the accretion disk and may significantly influence the inferred
neutron star masses.  This disk contribution is often non-negligible
(see Figs.\ 8, 9 of Rawls et al.\ 2011) and involves multiple
parameters. In order to assess the influence of this additional
information on our results, we reanalyze here the data of eclipsing
pulsars ignoring the ellipsoidal modulations.

For each of the six systems, we write the likelihood of the data as a
function of the neutron star mass $M_{\rm NS}$ as follows,
\begin{eqnarray}
P(\rm{data}|M_{\rm NS}) &=& C \int_{M_{\rm opt,min}}^{M_{\rm opt,max}} dM_{\rm opt}
\int_{\beta_{\rm min}}^{\beta_{\rm max}} d\beta
\int_{\Omega_{\rm opt,min}}^{\Omega_{\rm opt,max}} d\Omega_{\rm opt}
\int_{0}^{1} d(\cos i) \nonumber \\
&~& ~~~\exp\left[-\frac{(f_M-f_0)^2}{2\sigma_f^2}
-\frac{(K_{\rm opt}-K_0)^2}{2\sigma_K^2}
-\frac{(v_{\rm rot}\sin i-v_0)^2}{2\sigma_v^2}
-\frac{(\theta_e-\theta_0)^2}{2\sigma_\theta^2}
\right].
\label{eclipse}
\end{eqnarray}
Here $f_M = M_{\rm opt}^3 \sin^3i/(M_{\rm NS}+M_{\rm opt})^2$ is the
mass function of the binary, and $f_0$ and $\sigma_f$ are the measured
value of $f_M$ and its uncertainty (obtained from X-ray timing
observations).  Similarly, $K_0$, $\sigma_K$ are the measured value of
$K_{\rm opt}$ and its uncertainty, $v_0$, $\sigma_v$ are the measured
value of $v_{\rm rot}\sin i$ and its uncertainty, and $\theta_0$,
$\sigma_\theta$ are the measured semi-duration of the eclipse
$\theta_e$ and its uncertainty. All these measurements are listed in
Rawls et al.\ (2011) for the six systems of interest. Note that there
is no measurement of $v_{\rm rot} \sin i$ for Her X-1, which leads to
larger uncertainties in the mass determination of this source.

The quantity $\beta$ is equal to $R_{\rm opt}/R_L$, where $R_L$ is the
effective radius of the Roche lobe of the secondary.  For the
integration limits in equation (\ref{eclipse}), we choose $\beta_{\rm
min}=0$, $\beta_{\rm max}=1$, $\Omega_{\rm min}=0$, $\Omega_{\rm
max}=2\Omega_b$, where $\Omega_b=2\pi/P_b$ is the mean angular
velocity of the binary orbit, and a sufficiently generous range of
$M_{\rm opt}$.  As indicated in equation (\ref{eclipse}), we assume a
flat prior for each of the variables, though we have confirmed that
the results are not sensitive to this assumption. We calculate $P({\rm
data}|M_{\rm NS})$ for each of the six X-ray pulsars by computing the
integrals via a Monte Carlo method. We show in
Figure~\ref{fig:psr_slow} the resulting likelihoods of mass for each
of the six systems and summarize the results in Table~6.

We consider an eccentric orbit in the case of Vela X-1 using the value
of $e$ given in Rawls et al. (2011), and we assume a circular orbit
for the other five systems. Rawls et al. (2011) considered both
eccentric and circular orbits for 4U1538--52. However, the evidence
for eccentricity is not very strong. Moreover, the estimate of $K_{\rm
opt} =14.1\pm1.1 ~{\rm km\,s^{-1}}$ that they obtain for their
eccentric orbit solution appears to be anomalously low (it does not
fit the measured velocities very well --- see the lower left panel of
their Fig. 9). For this reason, we consider only a circular orbit for
4U1538--52, and we take $K_{\rm opt} = 21.8\pm3.8 ~{\rm km\,s^{-1}}$
as estimated by Rawls et al. (2011). (The fit to the velocity data
appears to be better in this case --- see the bottom right panel of
their Fig. 9.)


\begin{deluxetable}{lllllll}
\tabletypesize{\scriptsize}
\tablewidth{350pt}
\tablenum{6}
\tablecaption{Orbital Solutions for Eclipsing X-ray Pulsars}
\tablehead{
& \multicolumn{3}{c}{Rawls et al.\ (2011)\tablenotemark{a}} 
& \multicolumn{3}{c}{This Work}
\cr
 \colhead{Name} & \colhead{Mass} & \colhead{i} & $\beta$ & 
\colhead{Mass} & \colhead{i} & $\beta$ 
\cr 
\colhead{} & \colhead{$M_\odot$} & \colhead{deg} &  &
\colhead{$M_\odot$} & \colhead{deg} &  
}
\startdata
%
Vela X$-$1   & 1.770$\pm$0.083 & 78.8$\pm$1.2 & 1 
             & 1.70$\pm$0.13 & 86.3$\pm$2.6 & 0.99$\pm$0.01\\
4U 1538$-$52 & 0.996$\pm$0.101 & 76.8$\pm$6.7 & 0.88
             & 1.18$\pm$0.25 & 76.9$\pm$8.0 & 0.87$\pm$0.07\\
SMC X$-$1    & 1.037$\pm$0.085 & 68.5$\pm$5.2 & 0.95
             & 0.93$\pm$0.12 & 77.2$\pm$8.0 & 0.87$\pm$0.07\\
LMC X$-$4    & 1.285$\pm$0.051 & 67.0$\pm$1.9 & 0.95
             & 1.11$\pm$0.12 & 77.9$\pm$7.5 & 0.87$\pm$0.07\\
Cen X$-$3    & 1.486$\pm$0.082 & 66.7$\pm$2.4 & 1
             & 1.26$\pm$0.15 & 78.6$\pm$7.0 & 0.91$\pm$0.05\\
Her X$-$1    & 1.073$\pm$0.358 & $> 85.9$     & 1
             & 1.08$\pm$0.36 & 84.1$\pm$4.1 & 0.94$\pm$0.04 
\enddata
\scriptsize
\tablenotetext{a} {These values are taken from Table~4 of Rawls et al.\ (2011). 
}
\mbox{}
\end{deluxetable}

\begin{figure}
\centering
   \includegraphics[scale=0.75]{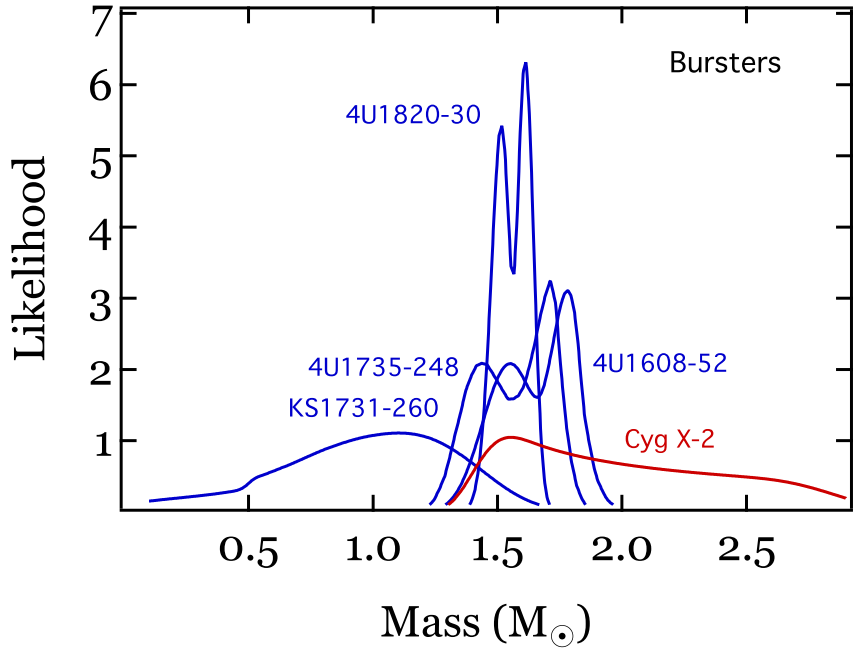}
\caption{The likelihoods $P_i({\rm data}\vert M_{\rm NS})$ for the accreting 
bursting neutron stars discussed as category V in the text. The likelihood 
for Cyg~X-2 was inferred using optical observations of its companion star. }
\label{fig:bursters} 
\mbox{}
\end{figure}

{\it (V) Accreting Bursting Neutron Stars.---\/} Spectroscopic
observations of a number of neutron stars during thermonuclear X-ray
bursts have led to measurements of their masses and radii. This
category includes 4U~1745$-$248 (\"Ozel et al.\ 2009), 4U~1608$-$52
(G\"uver et al.\ 2010a), 4U 1820$-$30 (G\"uver et al.\ 2010b), and
KS~1731$-$260 (\"Ozel et al.\ 2012). We convert the posterior
likelihood of mass and radius for each neutron star reported in
these analyses into a likelihood of mass by integrating over radius
as
\begin{equation}
P_i({\rm data} \vert M_{\rm NS}) = C_i \int dR_{\rm NS} P_i({\rm data} 
\vert M_{\rm NS}, R_{\rm NS}).
\end{equation}
The resulting likelihoods are shown in Figure~\ref{fig:bursters}. 

The mass of a fifth burster, Cyg X-2, has been measured in a different
manner using the optical observations of its companion star (Orosz \&
Kuulkers 1999). These observations yield the mass function of the
binary $f = 0.69 \pm 0.03~M_\odot$, a range of allowed mass ratios $0.3 < q <
0.38$, as well as a measurement of the binary inclination $i= 61^\circ
\pm 12^\circ$. Using this information, we calculate the posterior
likelihood of the mass of the neutron star according to (\"Ozel et
al.\ 2010a)
\begin{equation}
P_i({\rm data} \vert M_{\rm NS}) = C_i \int_{q_{\rm min}}^{q_{\rm max}} dq 
\int_{(\cos i)_{\rm min}}^1 \frac{d(\cos i)}{1-(\cos i)_{\rm min}}
\exp \left\{- \frac{[f_{0,i}-M_{\rm NS} \sin^3 i/(1+q)^2]^2} {2 \sigma_{f,i}^2}
- \frac{(i-i_0)^2}{2 \sigma_i^2}
\right\}
\label{eq:cygx2}
\end{equation}
and plot it in Figure~\ref{fig:bursters}.

\section{The Intrinsic Distribution of Neutron Star Masses}

The range of neutron star masses that can be produced in
astrophysically plausible scenarios is rather narrow, from $\sim
1.0~M_\odot$ to $\sim 2.5~M_\odot$. The posterior likelihood functions
for the neutron star mass measurements we report in the previous
section are indeed within this range. Moreover, earlier studies of the
neutron star masses indicated a narrowly peaked distribution (Finn
1994; Thorsett \& Chakrabarty 1999). For these reasons, we will model
the distribution of neutron star masses with a mean $M_0$ and a
dispersion $\sigma$, i.e.,
\begin{equation}
P(M_{\rm NS}; M_0,\sigma) = \frac{1}{\sqrt{2 \pi \sigma^2}} 
\exp\left[-\frac{(M_{\rm NS}-M_0)^2}{2 \sigma^2}\right].
\label{eq:gauss}
\end{equation} 
Our goal in this section is to obtain the most likely values for the
parameters $M_0$ and $\sigma$ of this distribution that are consistent
with the measurements. 

Using Bayes' theorem, we can write the probability $P(M_0,\sigma \vert
{\rm data})$ that measures the likelihood of the parameters of the
neutron star mass distribution as
\begin{equation}
P(M_0, \sigma \vert {\rm data}) = C P({\rm data} \vert M_0,\sigma) P(M_0) P(\sigma),
\label{eq:bayes}
\end{equation}
where $C$ is an appropriate normalization constant and $P(M_0)$ and
$P(\sigma)$ are the priors over the parameters of the mass
distribution. Hereafter, we will assume a flat prior over $M_0$
between 1~$M_\odot$ and $3~M_\odot$ as well as a flat distribution
over $\sigma$ between zero and $1~M_\odot$. 
We repeated the analysis using a logarithmic prior distribution, which had
indistiguishable effects on the final results. This is expected, given that
the inferred mass distributions are all very narrow and, therefore, are not
sensitive to any weak priors.

In equation~(\ref{eq:bayes}), the quantity $P({\rm data} \vert
M_0,\sigma)$ measures the posterior probability of having made a
particular set of observations for the ensemble of neutron stars given
the values of the parameters of the mass distribution. Under the 
assumption that each measurement is independent of all the others, 
we calculate this quantity using 
\begin{equation}
P({\rm data} \vert M_0,\sigma) = \prod_i \int dM_{\rm NS} 
P_i({\rm data} \vert M_{\rm NS}) P(M_{\rm NS}; M_0, \sigma).
\end{equation}
The only case where this assumption is not satisfied is for double
neutron stars in category IIa, for which the mass measurements of the
pulsar and the companion neutron star are not independent. For these 
three binary systems, we write instead 
\begin{eqnarray}
P({\rm data} \vert M_0,\sigma) &=& \prod_i^3 \int dM_{\rm tot} 
\exp\left[-\frac{(M_{\rm tot}-M_{\rm tot,0})^2}{2 
\sigma_{M_{\rm tot}}^2} \right] \nonumber \\
&\times&
\int dM_{\rm NS} 
P_i({\rm data} \vert M_{\rm NS}) P(M_{\rm NS}; M_0, \sigma) 
P(M_{\rm tot}-M_{\rm NS}; M_0, \sigma).
\end{eqnarray}

In Section~2, we divided the neutron star mass measurements not only
according to the technique by which these measurements were obtained
but also by the type of companion and the spin period of the neutron
star. We carried out the latter division in anticipation of the fact
that fast and slow pulsars are drawn from different parent
populations, i.e., from those which have and have not experienced
significant mass accretion phases. In the following, we derive the
parameters of the intrinsic mass distributions for each of those
populations separately. We then address the extent to which the
particular evolutionary channels followed by each type of neutron star
leaves a measurable imprint on the mass distribution.

\subsection{Neutron Stars at or Near their Birth Masses}

The low spin periods of a number of pulsars in our sample are
indicative of mild or even no recycling due to mass accretion. We,
therefore, consider the masses of neutron stars in this population
likely to be very near their birth values. This sample includes
categories Ia and IIa for double neutron stars, category IV for
accreting pulsars with primarily high-mass companions, as well as one
pulsar each in categories Ib (PSR~J1141$-$6545) and IIb
(PSR~B2303+46).

We first study the underlying mass distribution of the double neutron
stars. The mass measurements in these systems have by far the smallest
errors, which can dominate the parameter estimation of the mass
distribution of the total ensemble. Furthermore, these binaries have
followed a very particular and highly selective evolutionary path,
which may be evident in their mass distribution. We group the
remaining sources together as a second sample that consists of neutron
stars likely to be near their birth masses.

\subsubsection{Double Neutron Stars}

\begin{figure}
\centering
   \includegraphics[scale=0.75]{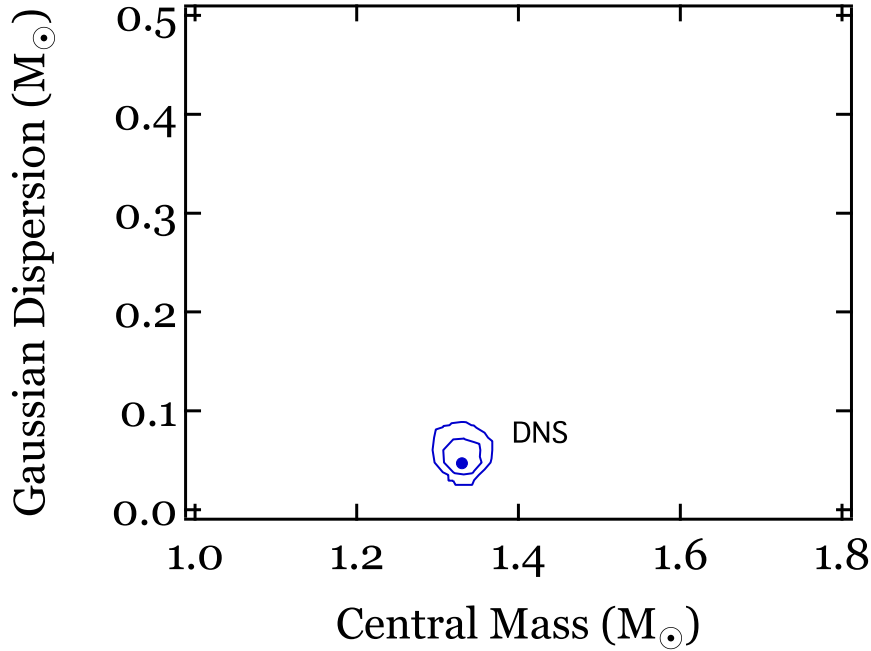}
\caption{The confidence contours over the parameters of a Gaussian 
distribution for the double neutron stars.}
\mbox{}
\label{fig:prob_dns} 
\end{figure}

Figure~\ref{fig:prob_dns} shows the 68\% and 95\% confidence contours
over the parameters of the intrinsic Gaussian distribution that is
consistent with the observed masses of nine double neutron stars. The
most likely value of the mean of the Gaussian distribution is
1.33~$M_\odot$ and that of the dispersion is 0.05~$M_\odot$. It is
evident from the figure that the uncertainties in the parameters of
the underlying distribution are very small: the 68\% errors are
0.03~$M_\odot$ in both parameters.

Our results are in agreement with the distribution reported by
Thorsett \& Chakrabarty (1999), who found a mean of 1.35~$M_\odot$ and
a dispersion of 0.04~$M_\odot$, and with the more recent results of
Kiziltan et al.\ (2010) for this category.

An interesting question we can address with the sample of double
neutron stars is whether the two members of each binary system are
drawn from the same underlying population. To study this, we divided
the sample into one of pulsars and one of the companions. Note that
for the double pulsar, we assigned the faster pulsar to the ``pulsar''
and the slower to the ``companion'' categories. Repeating the above
inference for these two subgroups individually, we found that the most
likely parameters of the mass distribution for the pulsars are $M_0 =
1.35~M_\odot$ and $\sigma = 0.05~M_\odot$, whereas for the companions
they are $M_0 = 1.32~M_\odot$ and $\sigma = 0.05~M_\odot$.  These
parameters are the same within the 68\% confidence ranges of each.

One further question about the double neutron star population concerns
the mass ratio $q$ in each binary. The mass ratios of the neutron
stars in two of the six binaries with well determined masses are
within one part in $\sim 5 \times 10^{-3}$ of unity. This is an order
of magnitude smaller than the most likely dispersion of the underlying
mass distribution. We explored whether the distribution of observed
mass ratios is consistent with the pulsar and the companion being
drawn independently from a Gaussian distribution with the parameters
we determined above.

\begin{figure}
\centering
   \includegraphics[scale=0.75]{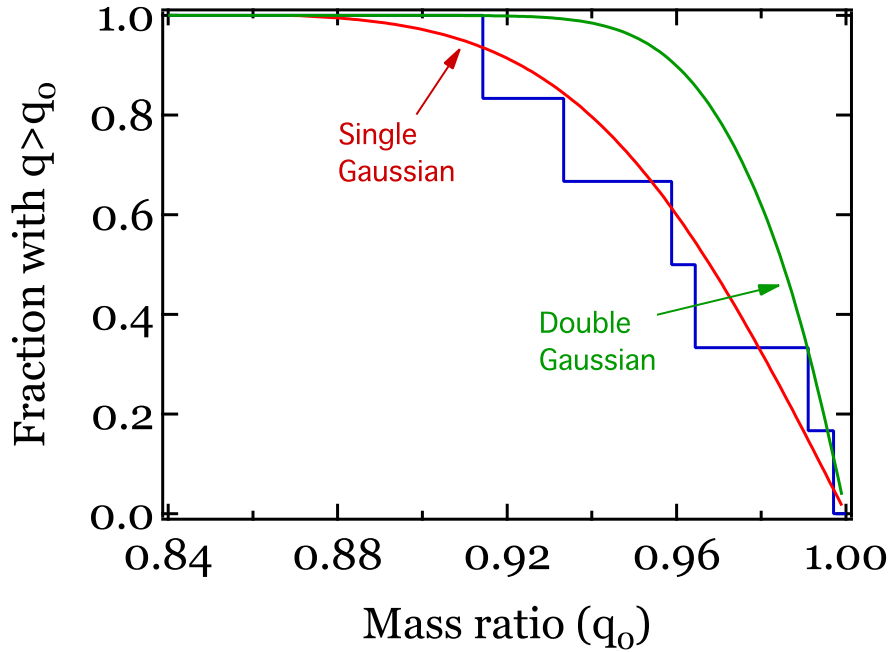}
\caption{The histogram shows the cumulative mass ratio distribution for 
the six double neutron stars with precise mass measurements. The red
line shows the predicted cumulative distribution for neutron star
pairs drawn independently from a single Gaussian distribution with a
central value and a dispersion equal to the most likely parameters
shown in Figure~\ref{fig:prob_dns}. The green line shows the predicted
cumulative distribution for neutron star pairs drawn independently
from the double Gaussian distribution suggested by Schwab et al.\
(2010). The observed distribution of mass ratios is in agreement with
a mass distribution represented by a single Gaussian. Note that, for
consistency, we show in this figure the mass ratio histogram generated
from the data used by Schwab et al.\ (2010).}
\mbox{}
\label{fig:dns_massratio} 
\end{figure}

The posterior likelihood of observing a binary with pulsar and
companion masses of $M_{\rm psr}$ and $M_{\rm c}$, respectively, is
given by
\begin{equation}
P(M_{\rm psr}, M_{\rm c}) dM_{\rm psr} dM_{\rm c} = C \exp 
\left[-\frac{(M_{\rm psr}-M_0)^2}{2\sigma^2}-\frac{(M_{\rm c}-M_0)^2}{2\sigma^2}\right]
dM_{\rm psr} dM_{\rm c}
\end{equation}
To convert this into a distribution over the mass ratio, we set 
$q \equiv \min(M_{\rm psr}/M_{\rm c}, M_{\rm c}/M_{\rm psr})$ and write 
\begin{eqnarray}
P(q)dq &=& \int_{M_{\rm psr}} P(M_{\rm psr},q) dM_{\rm psr} dq = 
\int_{M_{\rm psr}} P(M_{\rm psr},M_{\rm c}) 
\frac{dM_{\rm c}}{dq} dM_{\rm psr} dq \nonumber \\
       &=& \int_{M_{\rm psr}} \frac{2\;C}{M_{\rm psr}} 
\exp \left[-\frac{(M_{\rm psr}-M_0)^2}{2\sigma^2}-
\frac{(q\;M_{\rm psr}-M_0)^2}{2\sigma^2}\right]
dM_{\rm psr} dq\;.
\end{eqnarray}
In Figure~\ref{fig:dns_massratio}, we compare the cumulative
likelihood of the mass ratio
\begin{equation}
C(q>q_0)=\int_{q_0}^1 P(q)dq\;,
\end{equation}
calculated for the most likely values of the parameters of the
Gaussian, to the cumulative distribution of the observed mass ratios
of the 6 double neutron stars with well determined masses. The
similarity between the two distributions is striking and demonstrates
that the pulsar and the companion in each of the double neutron stars
are consistent with having been drawn independently from the same
narrow distribution of masses.

We also explored whether the observed distribution of mass ratios is
consistent with the predicted cumulative distribution for neutron star
pairs drawn independently from the double Gaussian distribution
suggested by Schwab et al.\ (2010; their eq.~(1)). The result is shown
as a green line in Figure~\ref{fig:dns_massratio}. The width of the
individual components in the bimodal distribution is significantly
narrower than the width of the single Gaussian that we infer here.
This leads to a larger fraction of double neutron stars with mass
ratios closer to unity for the bimodal distribution, which is not in
agreement with the observed sample.

\subsubsection{Accreting and Slow Pulsars}

The second subgroup consists of neutron stars accreting from high-mass
companions and slow pulsars, which are likely to be near their birth
masses. To infer the neutron star mass distribution, we will use both
the numerical results of Rawls et al.\ (2011) as well as our analytic
results discussed in \S 2, in which the information regarding
ellipsoidal variations in the lightcurves was not taken into account.

\begin{figure}
\centering
   \includegraphics[scale=0.75]{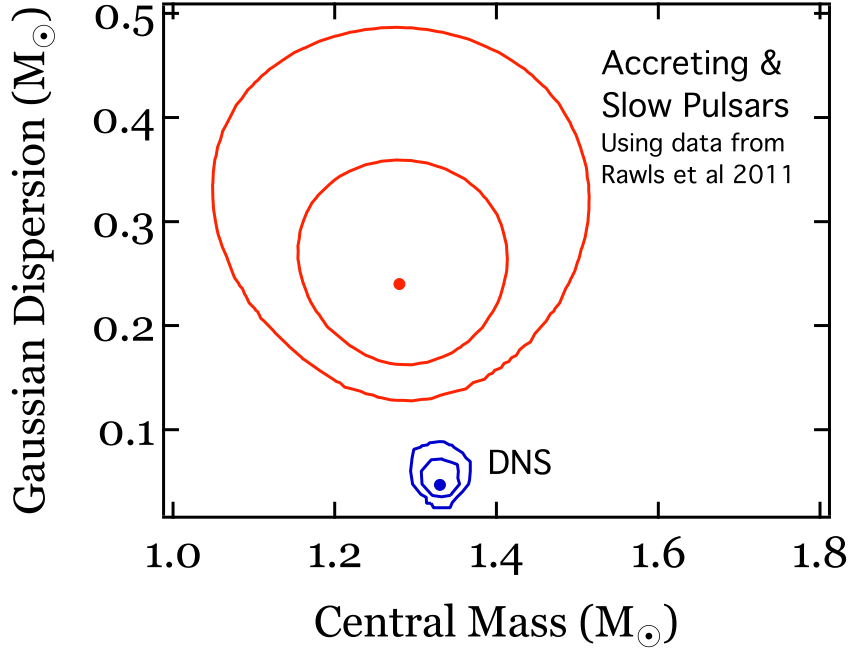}
\caption{The confidence contours over the parameters of a Gaussian 
distribution for the accreting and slow pulsars using the numerical
data from Rawls et al.\ (2011). The confidence contours for the double
neutron stars are also shown for comparison. Even though both
populations are thought to have masses near their birth masses, the
dispersion of double neutron star masses is significantly smaller than
that of the accreting and slow pulsars.}
\mbox{}
\label{fig:prob_slow_num} 
\end{figure}

In Figure~\ref{fig:prob_slow_num}, we show the 68\% and 95\%
confidence limits on the Gaussian parameters of the underlying mass
distribution for the accreting and slow pulsars, using the numerical
results of Rawls et al.\ (2011). For comparison, we overplot the
equivalent confidence contours for the double neutron stars. There is
a small but statistically insignificant shift in the central mass of
the Gaussian between the two populations. On the other hand, the
Gaussian dispersions between the two populations are different to a
high statistical significance. In other words, even though both of
these populations are believed to represent neutron stars near their
birth masses, the double neutron stars are drawn from a significantly
narrower distribution of masses. The most likely values of the central
mass and dispersion for the accreting and slow pulsars are
1.28~$M_\odot$ and 0.24~$M_\odot$, respectively.

\begin{figure}
\centering
   \includegraphics[scale=0.75]{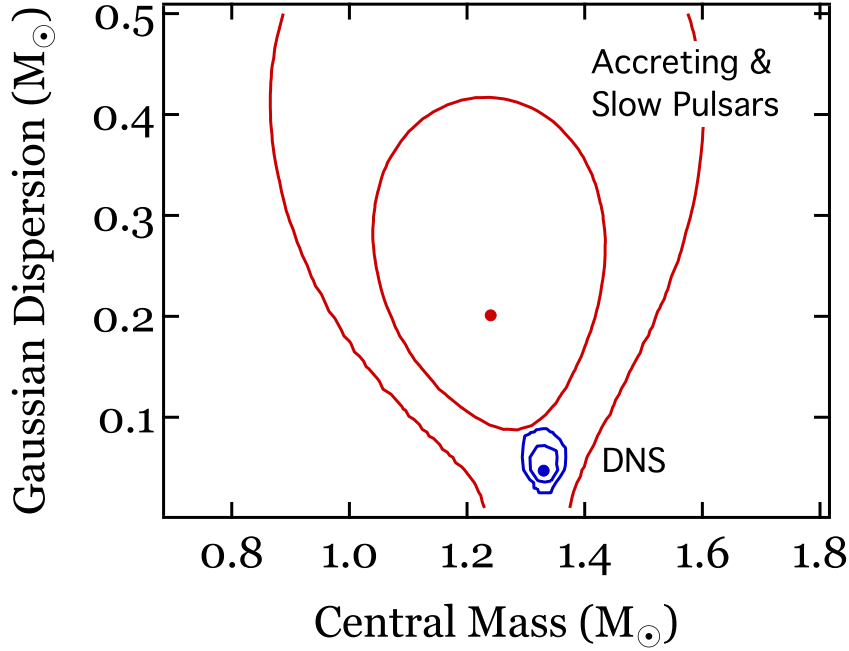}
\caption{The confidence contours over the parameters of a Gaussian 
distribution for the accreting and slow pulsars using the analytic
mass measurements discussed in the text. The confidence contours for
the double neutron stars are also shown for comparison. }
\mbox{}
\label{fig:prob_slow_anlytc} 
\end{figure}

The Rawls et al.\ (2011) analysis depends on a synthesis of a large
number of spectroscopic and photometric measurements of the binaries
that are used to infer the binary parameters. Fitting these
observations, and especially taking into account the ellipsoidal
variations requires complex models of the shape of the companion star
and of the relative contribution of light from the accretion
disk. Moreover, the photometric observations of the ellipsoidal
variations typically are the lowest signal-to-noise components of the
mass measurements. In order to assess the possible influence of the
modeling of the ellipsoidal variations on our results, we also infer
the underlying mass distribution using our analytical posterior
probabilities discussed in \S 3. Figure~\ref{fig:prob_slow_anlytc}
shows the resulting 68\% and the 95\% confidence contours for the
parameters of the mass distribution. In this case, the most likely
values of the central mass and dispersion for the accreting and slow
pulsars are 1.24~$M_\odot$ and 0.20~$M_\odot$, respectively. Comparing
these to the distribution inferred from the numerical results of Rawls
et al.\ (2011), we see that the central mass remains unchanged, but
the dispersion becomes less constrained and is even statistically
consistent with that of double neutron stars.

\subsection{Recycled Neutron Stars}

\begin{figure}
\centering
   \includegraphics[scale=0.75]{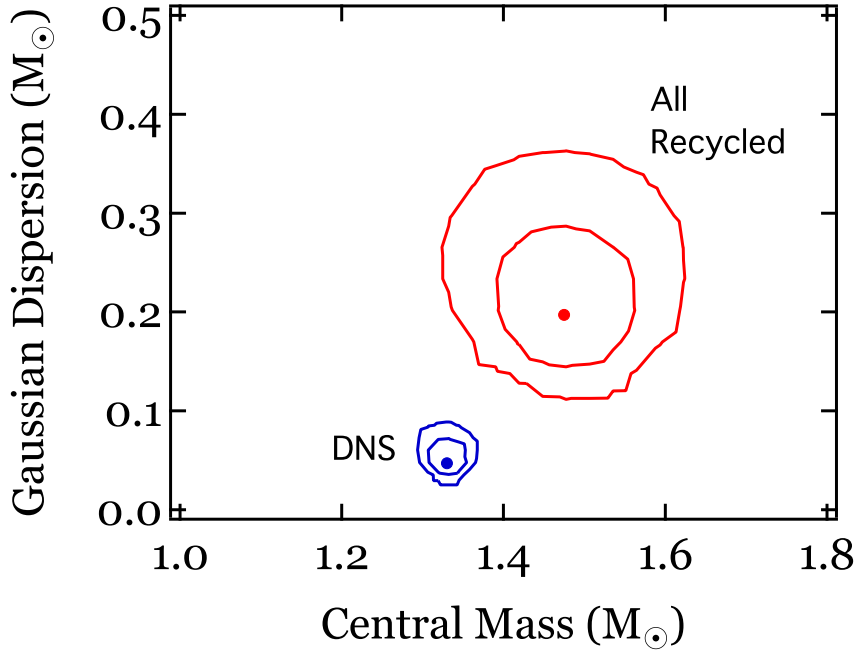}
\caption{The confidence contours over the parameters of a Gaussian 
distribution for the recycled neutron stars. The confidence contours
for the double neutron stars are also shown for comparison. As
expected, the recycled neutron stars have on average larger masses
than those in double neutron stars.}
\mbox{}
\label{fig:prob_recycl} 
\end{figure}

We now focus on the subgroup of neutron stars which have been recycled
through extended mass accretion. Neutron stars with white dwarf
companions, millisecond pulsars, as well as neutron stars in low-mass
X-ray binaries currently undergoing accretion (categories Ib, IIb,
III, and V) belong to this group.  Figure~\ref{fig:prob_recycl} shows
the confidence contours over the parameters of the Gaussian
distribution for the recycled neutron stars. The most likely value of
the central mass is 1.48~$M_\odot$ and of the dispersion is
0.20~$M_\odot$. Both the dispersion and the mean are similar to those
found by Kiziltan et al.\ (2010) within statistical uncertainties. The
uncertainties in the mean value quoted by Kiziltan et al.\ (2010),
however, are significantly smaller than those shown in
Figure~\ref{fig:prob_recycl} (c.f. Figure~3 of Kiziltan et al.\ 2010).

There are two main differences between our study and that of Kiziltan
et al. (2010) with regard to the recycled neutron star sample. First,
we make use of the detailed posterior likelihood for each mass
measurement, whereas Kiziltan et al.\ (2010) appear to have
approximated them with asymmetric Gaussians. Second, we include in our
sample neutron stars in low-mass X-ray binaries for which mass
measurements were performed mostly spectroscopically and typically
have larger uncertainties.

\begin{figure}
\centering
   \includegraphics[scale=0.75]{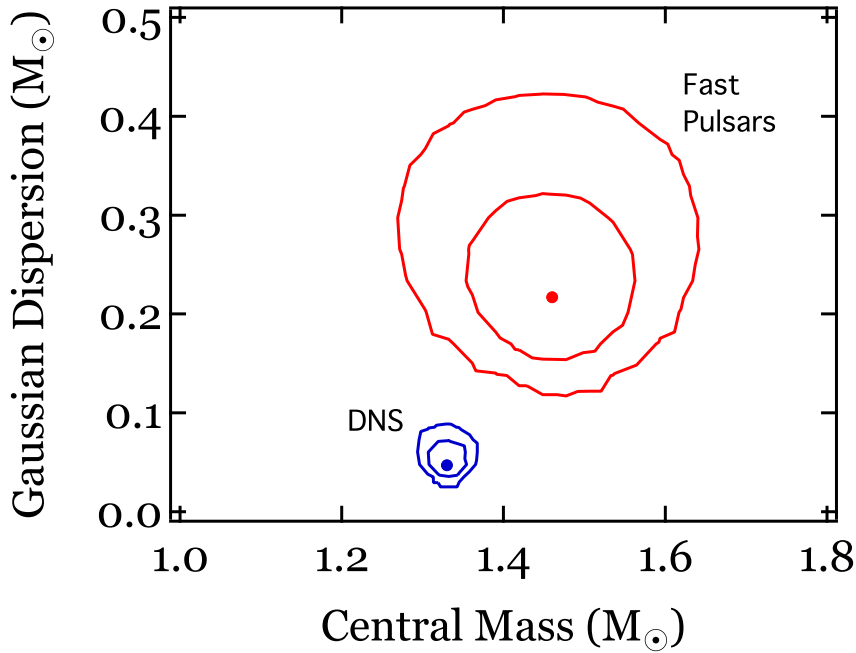}
\caption{The confidence contours over the parameters of a Gaussian 
distribution for a subgroup of the recycled neutron stars that
includes only pulsars in orbit around white dwarfs. Considering only
these sources with dynamical mass measurements does not alter the
results shown in Figure~\ref{fig:prob_recycl}.  }
\mbox{}
\label{fig:prob_fastpsr} 
\end{figure}

In order to assess the sensitivity of our results to the inclusion of
the accreting neutron stars, we repeat the inference of the mass
distribution parameters using only the radio pulsars in categories Ib
and IIb. In Figure~\ref{fig:prob_fastpsr} we show the resulting
confidence contours. The difference with the entire sample is minimal:
the most likely mean value and the dispersion are 1.46~$M_\odot$ and
0.21~$M_\odot$, respectively. We, therefore, attribute the small
difference with the Kiziltan et al.\ (2010) results to our handling of
the posterior likelihood distributions for each measurement.

\section{Discussion}

\begin{figure}
\centering
   \includegraphics[scale=0.75]{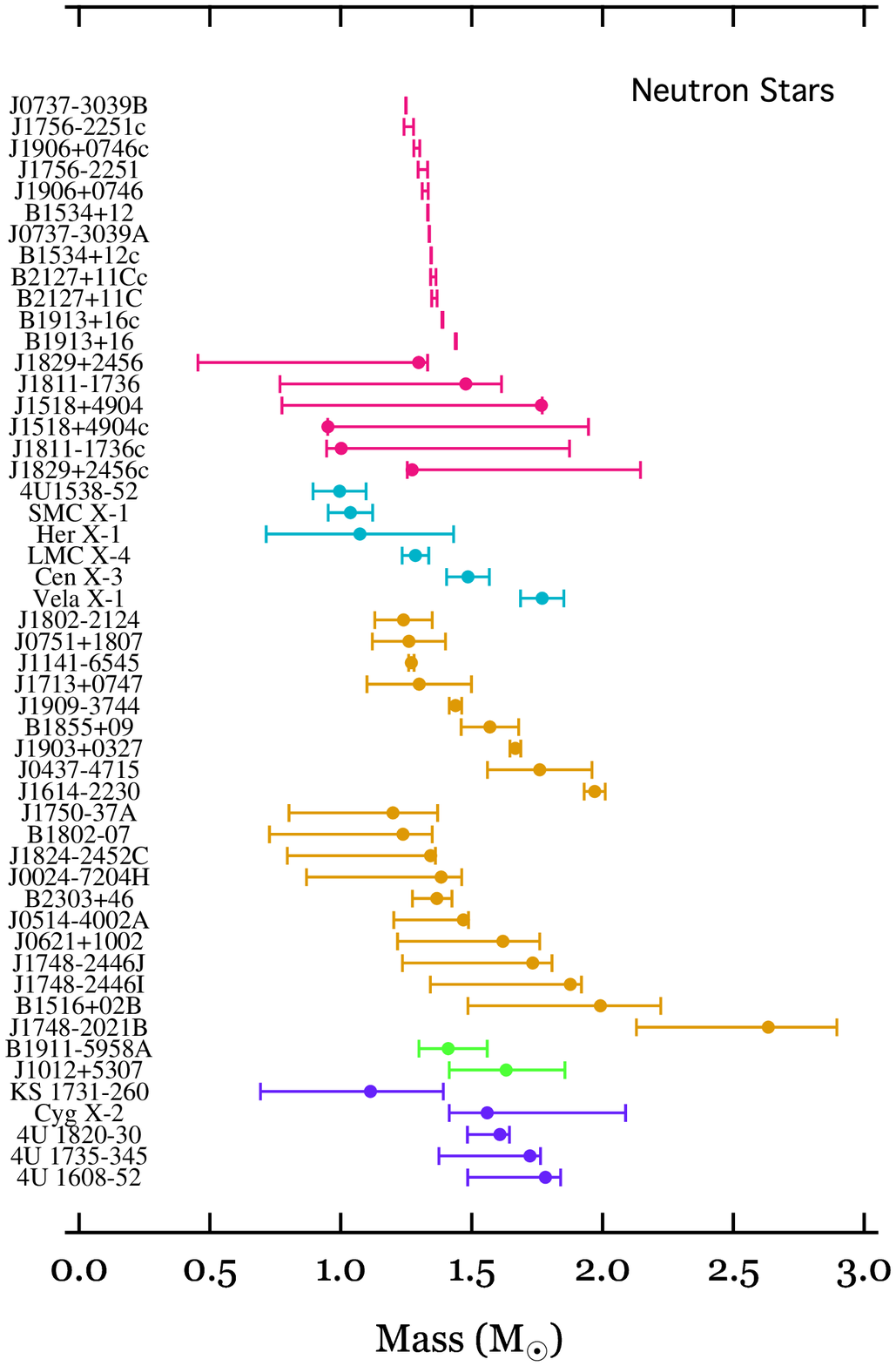}
\caption{The masses of neutron stars measured in double neutron stars 
(magenta; categories Ia and IIa), in eclipsing binaries with primarily
high mass companions (cyan; category IV; these are the numerical
values from Rawls et al.\ 2011 given in column~2 of Table~6), with
white dwarf companions (gold; categories Ib and IIb), with optical
observations of the white dwarf companions (green; category III), and
in accreting bursters (purple; category V). }
\mbox{}
\label{fig:nsmasses} 
\end{figure}

\begin{figure}
\centering
   \includegraphics[scale=0.75]{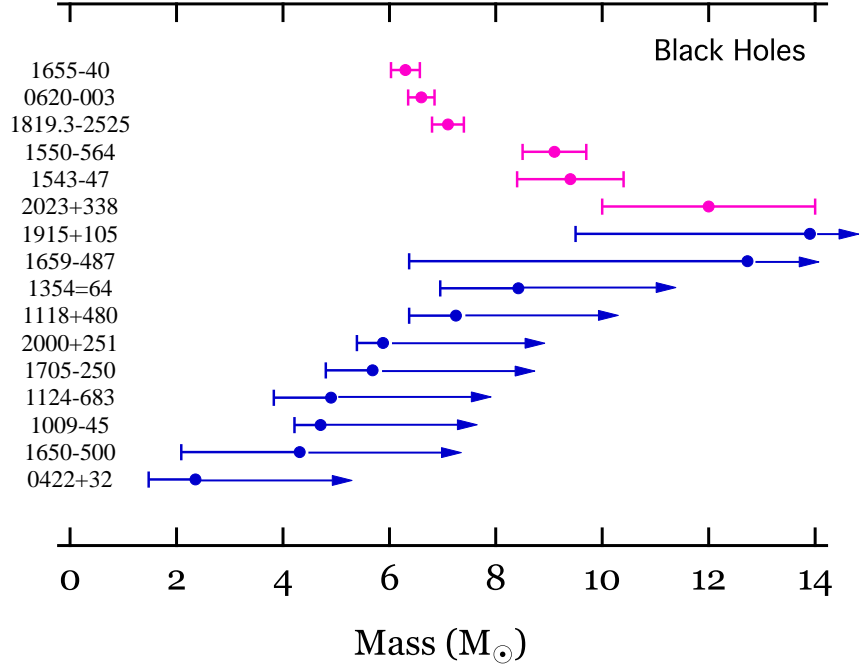}
\caption{The measured masses of Galactic black holes 
(after \"Ozel et al. 2010a).}
\label{fig:bhmasses} 
\mbox{}
\end{figure}

In this paper, we investigated the distribution of neutron star masses
in different types of binary systems and at different stages of
evolution based on currently available measurements. We summarize the
neutron star mass measurements and their uncertainties in each
subgroup in Figure~\ref{fig:nsmasses} and compare them to those of
black holes in Figure~\ref{fig:bhmasses} (compiled and analyzed in
\"Ozel et al.\ 2010a). In these figures, the error bars correspond 
to a 68\% confidence level calculated from the detailed likelihood
distribution presented for each subgroup of sources in \S 2.

\begin{figure}
\centering
   \includegraphics[scale=0.6]{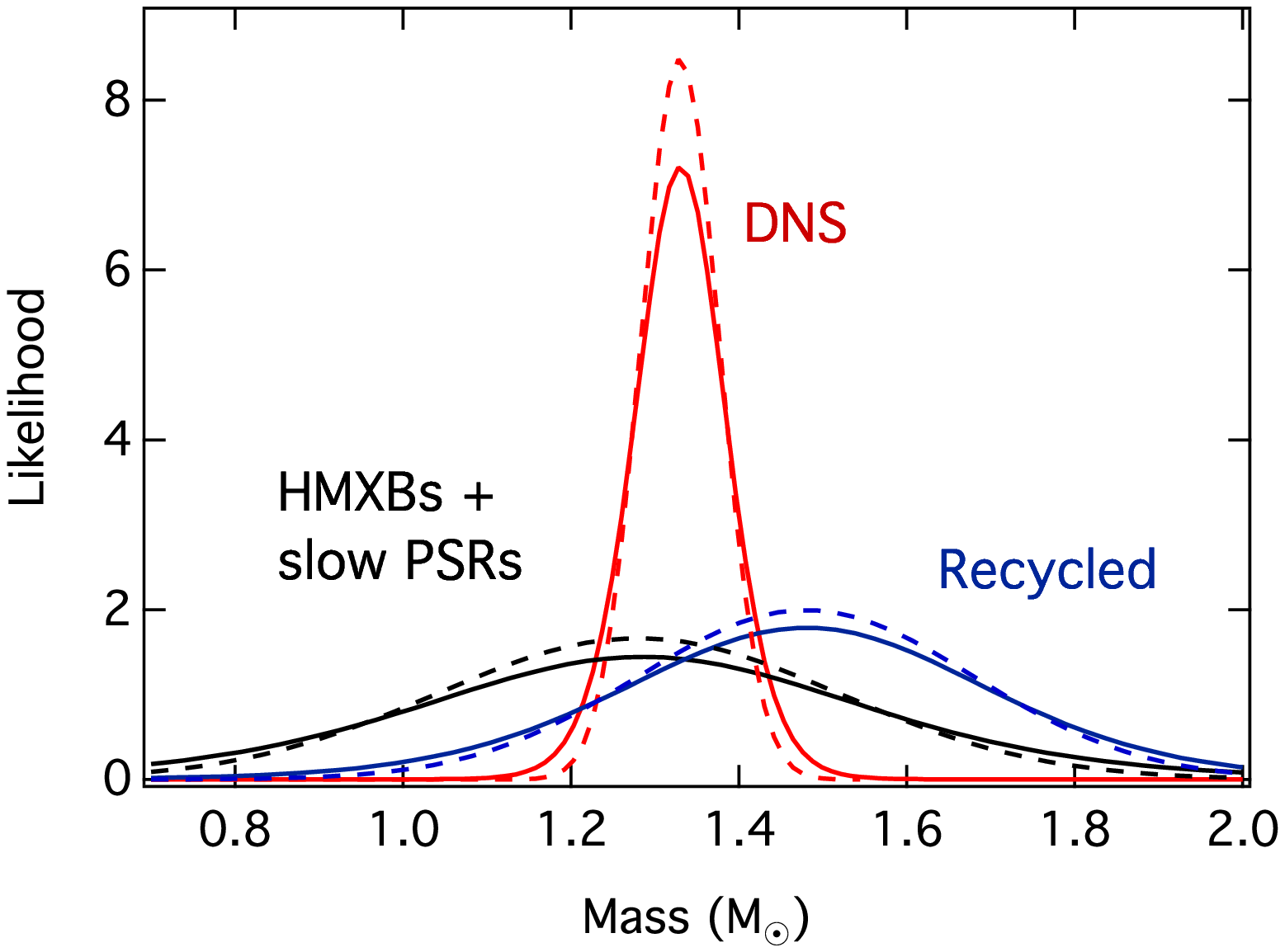}
   \includegraphics[scale=0.6]{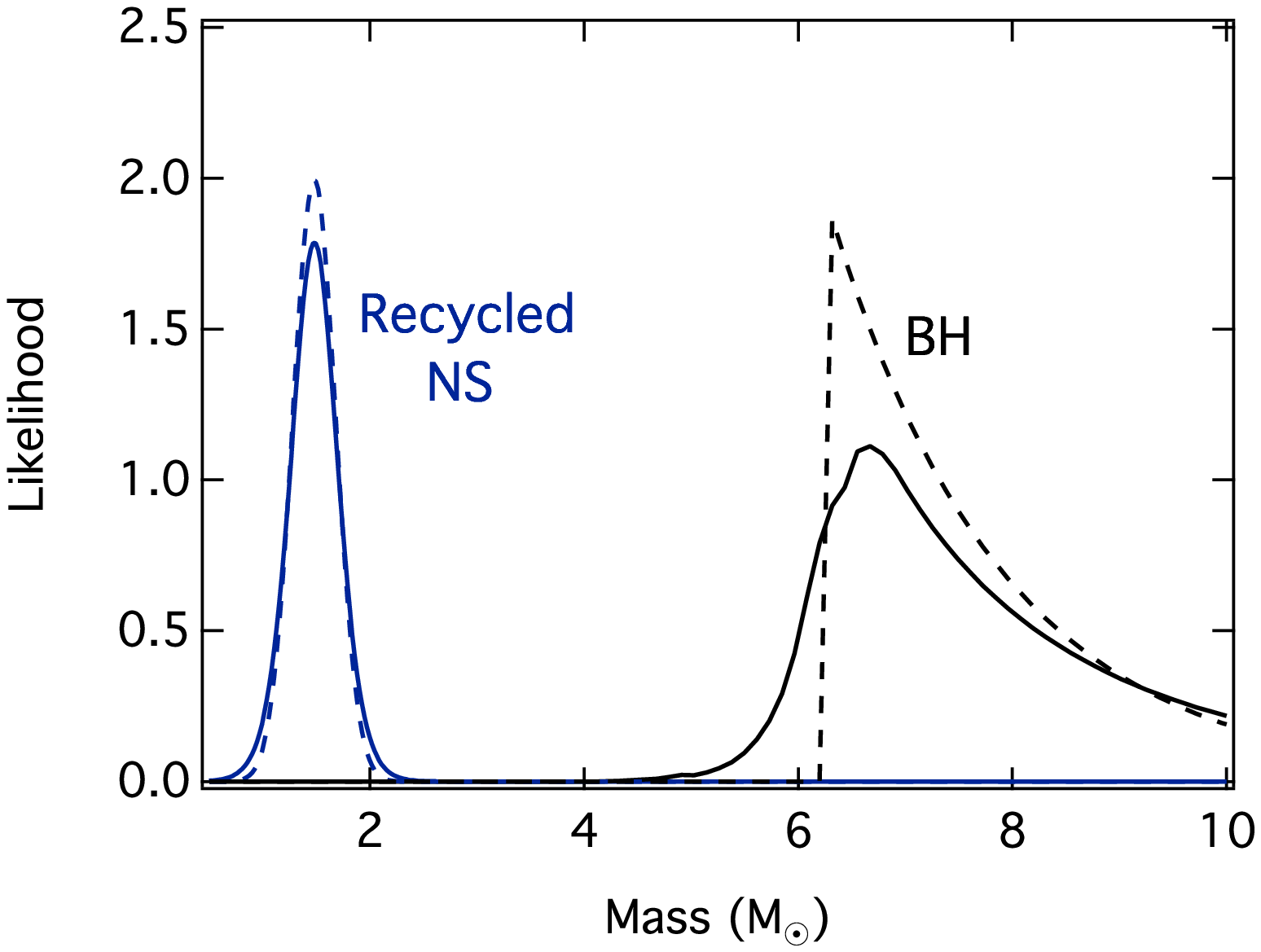}
\caption{The inferred mass distributions for the different populations 
of neutron stars (top) and black holes (bottom) discussed in the text.
The dashed lines correspond to the most likely values of the
parameters. For the different neutron star populations these are:
$M_0=1.33 M_\odot $ and $\sigma=0.05 M_\odot$ for the double neutron
stars, $M_0=1.28 M_\odot $ and $\sigma=0.24 M_\odot$ for the other
neutron stars near their birth masses, and $M_0=1.48 M_\odot $ and
$\sigma=0.20 M_\odot$ for the recycled neutron stars. For the case of
black holes, we used the exponential distribution with a low mass
cut-off at $M_{\rm c}=6.32 M_\odot$ and a scale of $M_{\rm scale}=1.61
M_\odot$ obtained in \"Ozel et al.\ (2010a). The solid lines represent
the weighted mass distributions for each population, for which
appropriate fitting formulae are given in the Appendix. The
distributions for the case of black holes have been scaled up by a
factor of three for clarity. }
\mbox{}
\label{fig:weighted} 
\end{figure}

In the top panel of Figure~\ref{fig:weighted}, we show the inferred
mass distributions of the various neutron star populations discussed
in Section~2. For each population, we present two different
distributions. The dashed lines correspond to the most likely
parameters of the underlying distributions inferred in Section~3. Each
solid line represents the weighted distribution over the central mass
and dispersion for each population. We compute this weighted
distribution as
\begin{equation}
P_w(M_{\rm NS})=\int dM_0 \int d\sigma P(M_{\rm NS}; M_0,\sigma)
P(M_0,\sigma \vert {\rm data}), 
\end{equation}
where $P(M_{\rm NS}; M_0,\sigma)$ and $P(M_0,\sigma \vert {\rm data})$
is given by equations~\ref{eq:gauss} and \ref{eq:bayes},
respectively. In the Appendix, we provide approximate analytic fitting
formulae for these weighted distributions for each population.

In the bottom panel of Figure~\ref{fig:weighted}, we compare the
inferred mass distribution for recycled neutron stars to that of black
holes reported in \"Ozel et al.\ (2010a). For the latter, we use the
exponential model with a lower mass cutoff given by
\begin{equation}
P(M_{\rm BH}; M_{\rm scale}, M_{\rm c}) = 
\frac{\exp(M_{\rm c}/M_{\rm scale})}{M_{\rm scale}}
\times\left\{
\begin{array}{ll}
\exp(-M_{\rm BH}/M_{\rm scale}), & M_{\rm BH}>M_{\rm c} \\
0, & M_{\rm BH}\leq M_{\rm c}
\end{array}
\right..
\end{equation}
The most likely values for the parameters of this distribution are
$M_{\rm scale}=1.61 M_\odot$ and $M_{\rm c} = 6.32 M_\odot$. In the
same panel, we also include the appropriate weighted distribution for
the black holes, where we carried out the integration over the
posterior likelihood of the parameters $M_{\rm scale}$ and $M_{\rm
c}$; we provide an analytic fitting formula for the weighted
distribution in the Appendix. This panel highlights the substantial
mass gap that exists between the black hole population and even the
heaviest neutron star population (see the discussion in
\"Ozel et al.\ 2010a and Farr et al.\ 2011).

Within the neutron star population, it is evident from these figures
that the mass distribution of double neutron star systems is different
than those observed in other binary systems, which include both
neutron stars near their birth masses as well as neutron stars that
experienced significant accretion episodes. Indeed, the most likely
values of the mean mass and the dispersion we derived for these
populations using the Bayesian inference technique discussed in \S 3
are $1.33 \pm 0.05~M_\odot$ for double neutron stars, in contrast to
$1.28 \pm 0.24~M_\odot$ for other neutron stars near their birth
masses and $1.48 \pm 0.20~M_\odot$ for recycled neutron stars. Note
that the uncertainties in both the mean mass and the dispersion for
all of these subgroups are shown in Figures~9-12.

\begin{figure}
\centering
   \includegraphics[scale=0.75]{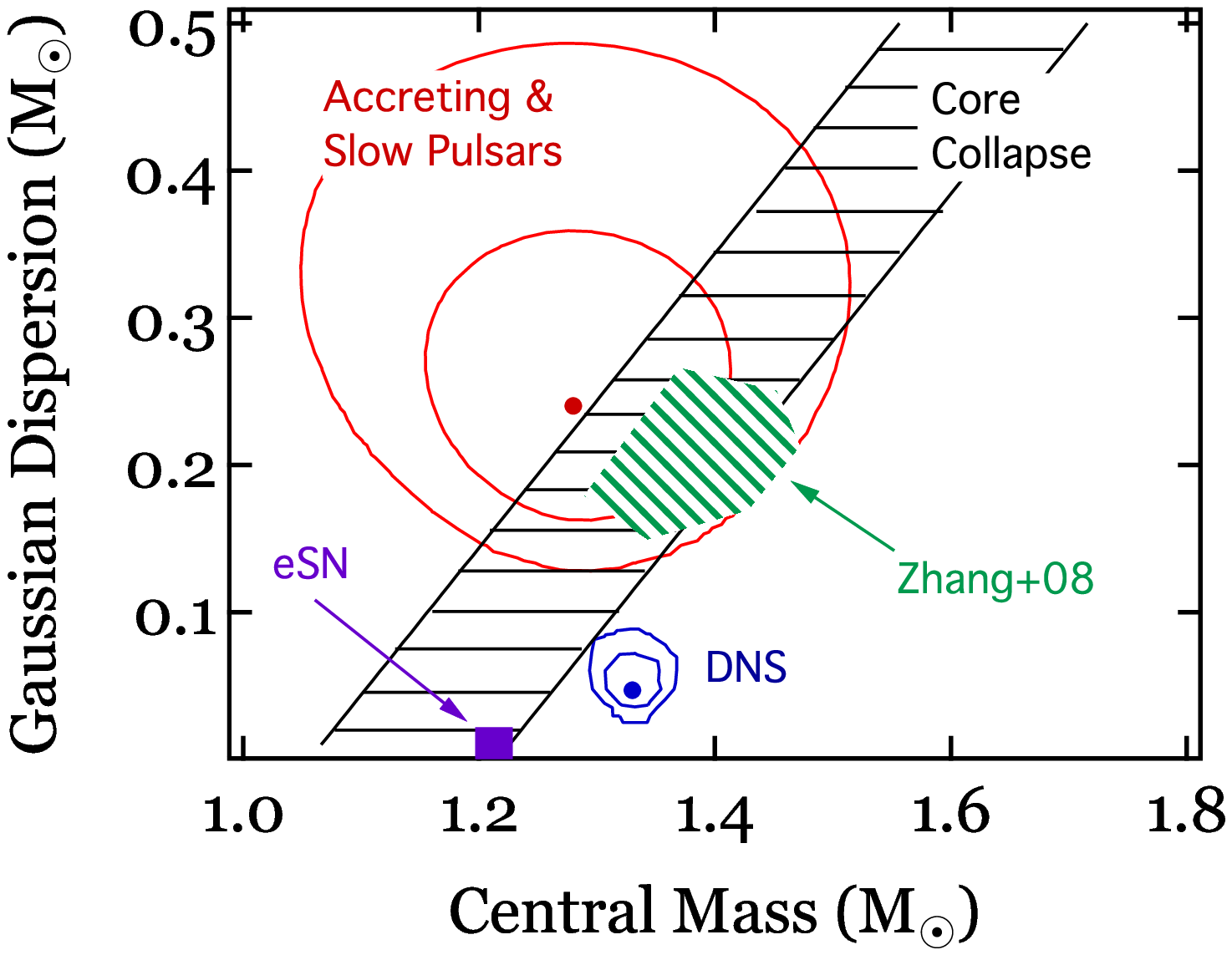}
\caption{The distribution of neutron star masses at birth expected from 
theoretical calculations, compared to the observed mass distribution
of neutron stars that have not accreted significant amounts of mass
(labeled accreting and slow pulsars) as well as to that of double
neutron stars. The parameters of the distribution of the former
subgroup are consistent with expectations from core-collapse supernova
and fallback, while the observed double neutron star distribution is 
significantly narrower than what is expected.} 
\mbox{}
\label{fig:nsmass_theory} 
\end{figure}

The narrowness of the mass distribution of double neutron stars is
difficult to account for within the current understanding of neutron
star formation mechanisms. One possible way to generate a narrow
distribution is via electron capture supernovae in ONeMg white dwarfs.
The onset of such a supernova occurs at a particular density
threshold, which corresponds to a pre-collapse mass of the white dwarf
in the narrow range $1.36-1.38~M_\odot$ for different temperatures and
compositions (Podsiadlowski et al.\ 2005). Taking into account a
binding mass given by the approximate formula (Lattimer \& Yahil 1989)
\begin{equation}
E_B = 0.084 \left(\frac{M}{M_\odot}\right)^2 M_\odot
\end{equation}
the gravitational masses of the outcomes of electron capture
supernovae become $1.2-1.22~M_\odot$. This range of masses is compared
to the parameters of the underlying distributions of double neutron
stars as well as of the other neutron stars near their birth masses in
Figure~\ref{fig:nsmass_theory}. Even though the electron capture
supernovae are capable of producing a narrow range of neutron star
masses, the mean of the expected distribution is inconsistent with
that of double neutron stars to a high confidence level.

In the case of core collapse supernovae, the mean mass of the neutron
stars produced in the absence of fallback is also expected to be
significantly smaller than that inferred from the double neutron
stars. We can estimate this mass assuming that the core of the
pre-supernova star collapses when it reaches its Chandrasekhar limit.
Considering an electron fraction of $Y_e = 0.42-0.48$, which is
appropriate for the cores of presupernova stars (Timmes et al.\ 1996), 
the Chandrasekhar mass 
\begin{equation}
M_{\rm Ch} = 5.83 Y^2_e
\end{equation}
falls in the range $1.15-1.34~M_\odot$. Taking into account the
gravitational binding energy, the expected range of birth masses for
neutron stars from core-collapse supernovae is $1.06-1.22~M_\odot$.

Fallback of matter during and immediately following the supernova
explosion can naturally lead to neutron stars more massive than the
cores of the progenitor stars. At the same time, the stochastic nature
of fallback necessarily leads to an increased dispersion of neutron
star masses. In Figure~\ref{fig:nsmass_theory}, we show the evolution
of the expected dispersion with central mass assuming that a fallback
of baryonic mass $\Delta M_f$ introduces a dispersion of the baryonic
mass of the neutron star of the same magnitude. (Note that in
Fig.~\ref{fig:nsmass_theory}, we plot the corresponding gravitational
mass for the neutron stars.) This simple analytical estimate is in
agreement with the detailed numerical calculations of Zhang et al.\
(2008), which are also shown in the figure. The green hatched region
outlines the results for different compositions, explosion energies,
and locations of the pistons for an assumed maximum neutron star mass
of $2~M_\odot$. Allowing sufficient fallback to account for the mean
value of the double neutron star masses introduces a dispersion that
is significantly larger than the observed one. In contrast, the
inferred mean and dispersion of the mass distribution of other neutron
stars believed to be near their birth masses (labeled accreting and
slow pulsars) are in agreement with theoretical expectations of core
collapse supernovae.

Considering a bimodal mass distribution for the double neutron stars,
as in Schwab et al.\ (2010), aggravates these challenges. First, as we
showed in \S 3, the cumulative likelihood of the mass ratio for such
a distribution does not agree with the cumulative distribution of the
observed mass ratios for the double neutron stars. Second, the
dispersion in the two components becomes even smaller, $\simeq
0.025~M_\odot$, making the higher mass component even less consistent
with the expectations of the core collapse supernovae. All of these
arguments lead to the conclusion that the mass distribution of double
neutron stars is peculiar and perhaps related to the particular
evolutionary history that leads to their formation.

The masses of the population of recycled neutron stars, which include
fast pulsars with white dwarf companions as well as accreting
bursters, are consistent with them having undergone extended periods
of accretion. On average, recycled neutron stars are more massive by
$\approx 0.2 M_\odot$ compared to other accreting and slow
pulsars. Such a mass increase is more than adequate to recycle these
pulsars to millisecond periods. Indeed, assuming that the mass is
transferred onto the neutron star via an accretion disk that is
magnetically truncated at the corotation radius
\begin{equation}
R_c = \left(\frac{GM}{4 \pi^2 \nu^2_{\rm s}}\right)^{1/3}, 
\end{equation}
where $\nu_{\rm s}$ is the spin frequency of the neutron star, 
the angular momentum transferred per unit mass is 
\begin{equation}
l= (G M R_c)^{1/2} = \left(\frac{G^2 M^2}{2 \pi \nu_{\rm s}}\right)^{1/3}.
\end{equation}
After accreting mass $\Delta M$, the neutron star acquires an angular 
momentum $\Delta M \cdot l$. Equating this to the spin angular momentum of the 
recycled pulsar $L=2 \pi I \nu_{\rm s}$, where $I$ is its moment of inertia, 
allows us to calculate the mass required to spin up the pulsar as 
\begin{equation}
\Delta M = I (GM)^{-2/3} (2 \pi \nu_{\rm s})^{4/3} 
= 0.034 \left(\frac{\nu_{\rm s}}{300~{\rm Hz}}\right)^{4/3}
\left(\frac{M}{1.48~M_\odot}\right)^{-2/3}
\left(\frac{I}{10^{45}~{\rm g}~{\rm cm}^2}\right)\;M_\odot. 
\end{equation}

It is interesting that the most likely value of the mean mass of the
recycled pulsars is significantly smaller than the $2~M_\odot$ lower
bound on the maximum mass of a neutron star (Demorest et al.\ 2010;
\"Ozel et al.\ 2010b) as well as the average mass of recycled
neutron stars predicted by population synthesis studies (e.g., Pfahl,
Rappaport, \& Podsiadlowski 2003; Lin et al.\ 2011). This conclusion
can be used to refine models of low-mass X-ray binary evolution.
Furthermore, our analysis shows that a very small fraction of neutron
stars reach masses comparable to the maximum possible neutron star
mass and collapse into black holes. Therefore, this channel does not
contribute significantly to a putative but still undetected population
of low-mass black holes in the Galaxy (see the discussion in \"Ozel et
al.\ 2010a).

\acknowledgments

F.\"O., D.P., and R.N. thank the Institute of Astronomy, where a large
fraction of this work was carried out, for their hospitality. We thank
Scott Ransom for useful discussions as well as Ingrid Stairs, Michael
Kramer, and Matthew Bailes for detailed comments on the manuscript. We
thank Laura Kasian for sharing unpublished data. We gratefully
acknowledge support from NSF grant AST-1108753, NASA ADAP grant
NNX10AE89G, NSF CAREER award NSF 0746549, and Chandra Theory grant
TMO-11003X for this work.

\appendix
\section{Fitting Formulae for Weighted Mass Distributions of Neutron Stars 
and Black Holes}

The weighted distributions of all of the neutron star populations,
shown as solid lines in Figure~\ref{fig:weighted}, are well
approximated by Gaussian functions, with a mean and dispersion of $M_0
= 1.33 M_\odot$ and $\sigma=0.072 M_\odot$ for the double neutron
stars, $M_0 = 1.28 M_\odot$ and $\sigma=0.28 M_\odot$ for other
neutron stars near their birth mass, and $M_0 = 1.48 M_\odot$ and
$\sigma=0.22 M_\odot$ for recycled neutron stars.

We also obtained a fitting formula for the normalized weighted mass
distribution of black holes (solid line in the bottom panel of
Fig.~\ref{fig:weighted}) for $M_{\rm BH}> 5 M_\odot$ that approximates the
numerical result to within $3\%$:
\begin{equation}
P(M_{\rm BH})=\left\{A(M_{\rm BH})^n+
\left[B(M_{\rm BH})^{-n}+C(M_{\rm BH})^{-n}
 \right]^{-1}\right\}^{1/n}\;,
\end{equation}
where 
\begin{eqnarray}
A(M_{\rm BH})&=&4.367-1.7294 M_{\rm BH}+0.1713 M_{\rm BH}^2\nonumber\\
B(M_{\rm BH})&=&14.24\exp(-0.542 M_{\rm BH})\nonumber\\
C(M_{\rm BH})&=&3.322\exp(-0.386 M_{\rm BH})\nonumber\\
n&=&-10.0\;.
\end{eqnarray}

\end{document}